\documentclass[a4paper,fleqn,usenatbib]{mnras}

\usepackage{txfonts}

\usepackage[T1]{fontenc}
\usepackage{ae,aecompl}

\usepackage{graphicx}
\usepackage{amssymb}
\usepackage{subfigure}

\title[Constraints on star formation in NGC\,2264]{Constraints on star formation in NGC\,2264}

\author[R.~J.~Parker \& C.~Schoettler]{Richard  J. Parker\thanks{E-mail: R.Parker@sheffield.ac.uk}\thanks{Royal Society Dorothy Hodgkin Fellow} and Christina Schoettler\vspace*{0.1cm}\\
Department of Physics and Astronomy, The University of Sheffield, Hicks Building, Hounsfield Road, Sheffield, S3 7RH, UK}

\begin{document}

\date{}
                             
\pagerange{\pageref{firstpage}--\pageref{lastpage}} \pubyear{2021}

\maketitle

\label{firstpage}

\begin{abstract}
We quantify the spatial distribution of stars for two subclusters centred around the massive/intermediate mass stars S~Mon and IRS\,1/2 in the NGC\,2264 star-forming region. We find that both subclusters are have neither a substructured, nor a centrally concentrated distribution according to the $\mathcal{Q}$-parameter. Neither subcluster displays mass segregation according to the $\Lambda_{\rm MSR}$ ratio, but the most massive stars in IRS\,1/2 have higher relative surface densities according to the $\Sigma_{\rm LDR}$ ratio. We then compare these quantities to the results of $N$-body simulations to constrain the initial conditions of NGC\,2264, which are consistent with having been dense ($\tilde{\rho} \sim 10^4$\,M$_\odot$\,pc$^{-3}$), highly substructured and subvirial. These initial conditions were also derived from a separate analysis of the runaway and walkaway stars in the region, and indicate that star-forming regions within 1\,kpc of the Sun likely have a broad range of initial stellar densities. In the case of NGC\,2264, its initial stellar density could have been high enough to cause the destruction or truncation of protoplanetary discs and fledgling planetary systems due to dynamical encounters between stars in the early stages of its evolution. 

\end{abstract}


\section{Introduction}

Stars form from the collapse and fragmentation of Giant Molecular Clouds, which usually results in groups or clusters of stars of a similar age and chemical composition \citep{Lada03}. These groups tend to exhibit spatial and kinematic substructure at young ages \citep[e.g.][]{Gomez93,Gouliermis14,Foster15,Buckner19} as well as stellar densities that are between $10 - 10^5$ times as dense as the Sun's current location in the Galaxy \citep{Korchagin03}. 

Because of their high densities, these star-forming regions tend to evolve on very short timescales ($\sim$1\,Myr) and so establishing the initial conditions (e.g.\,\,birth density, velocity distribution) can be problematic \citep{Parker14e}. The most effective way of establishing the initial conditions of a star-forming region is to ``reverse engineer" the present-day observed spatial and kinematic distributions, i.e.\,\,compare the current distributions to numerical simulations of the formation and evolution of star-forming regions.

In order to do this, an assumption has to be made about the universality of star-formation, i.e. a property must be fixed that is thought to be the general outcome of all star formation \citep[such as the Initial Mass Function (IMF),][]{Bastian10}. Some authors have assumed the primordial binary star population -- like the IMF -- is invariant, and therefore the amount of binary destruction places limits on the amount of dynamical evolution \citep{Kroupa95a,Marks11}. 

However, the observed binary populations suffer from high levels of incompleteness \citep{Duchene13b}, and furthermore, the narrow range of observed binaries in star-forming regions straddle the `Hard-Soft' boundary \citep{Heggie75,Hills75a,Hills75b}, making it difficult to ascertain whether an observed binary population will have undergone dynamical processing, even if situated in a dense star-forming region \citep{Parker12b}. 

As an alternative to binary stars, in previous work \citep{Parker14b,Wright14,Parker17a} we have attempted to constrain the initial conditions of star formation by assuming that each star-forming event results in a spatially and kinematically substructured distribution (where the level of substructure is allowed to vary), and use this substructure as a dynamical clock to compare to observations. 

This earlier work focused exclusively on the spatial information, such as the overall structure \citep{Cartwright04,Cartwright09b} and the amount of mass segregation \citep{Allison09a} or relative surface densities \citep{Maschberger11,Parker14b}. The \emph{Gaia} satellite has facilitated a vast improvement in these comparisons by enabling the proper motions of runaway stars to be used as dynamical tracers of the regions' past evolution \citep{Schoettler19,Farias20,Schoettler20}. 

Recently, \citet{Schoettler20} used runaway ($>30$\,km\,s$^{-1}$ and walkaway ($\gtrsim 10$\,km\,s$^{-1}$) stars to confirm earlier structural analyses \citep{Allison09a,Allison10,Allison11,Parker14b} that constrained the initial conditions of the Orion Nebula Cluster as being dense, substructured and subvirial. Using information from the proper motions of runaways and walkaways has the advantage over studies that quantify the radial velocity dispersion in that no correction for binary stars need be applied \citep{Gieles10,Cottaar12b}. 

In \citet{Schoettler21b} we applied this analysis to NGC\,2264, which unlike the ONC is not a smooth centrally concentrated cluster. Indeed, NGC\,2264 is referred to as the `Christmas Tree cluster', due to its unusual spatial distribution. NGC\,2264 has had at least two temporally distinct episodes of star formation, but \citet{Schoettler21b} was able to pinpoint the initial conditions of these two episodes and -- like the ONC -- found them to be consistent with dense ($10^4$\,M$_\odot$\,pc$^{-3}$), substructured and subvirial star-formation. 

In this paper, we follow-up the analysis of NGC\,2264 in \citet{Schoettler21b} to determine whether the spatial distribution of the stars in this region is also consistent with the analysis of the runaway and walkaway stars. This paper is organised as follows. In Section~\ref{data} we provide a description of the dataset, in Section~\ref{methods} we describe the methods used to quantify the spatial distributions of stars, as well as describing the set-up of the $N$-body simulations we compare the data to. In Section~\ref{results} we present the results, we provide a discussion in Section~\ref{discuss} and we conclude in Section~\ref{conclude}.

\section{Observational data}
\label{data}

We construct a mass census of stars that belong to NGC\,2264 based on identified members from different surveys. The member surveys we use contain a larger number of stars than those we include into our mass census as some of these stars lack information on their stellar mass.

For the mass census, we identify members from the work of \citet{Venuti17,Venuti18,Venuti19}, 
who mainly focused on the low- to intermediate mass members of the cluster. Their work is mainly based on photometric observations from the ``Coordinated Synoptic Investigation of NGC 2264'' (CSI) 2264 project \citep{Cody14} in combination with data from other photometric and spectroscopic observational campaigns \citep[e.g. SDSS, Gaia-ESO, IPHAS, Pan-STARRS1,][]{Abazajian09,Randich13,Barentsen14,Flewelling20}. 

We find further members in \citet{Dahm07} who used X-ray observations; \cite{Jackson20} who used data from the Gaia-ESO campaign; \citet{Maiz19} and \citet{CantatGaudin20} who both used \textit{Gaia} DR2 for their member identification. Where information about the membership probability was given, we use stars with a probability of 50 per cent or above as part of our census. As noted in \citet{Schoettler21b}, while the members of NGC\,2264 are fairly well constrained on-sky, they are far less constrained in their distance measurement. While the majority of the members are found around at a distance of around 700--800 pc, there are also identified members in several of the literature sources we use that show distances far above and below this value range. We do not exclude any of these members in our analysis, as a comparison of the \textit{Gaia} DR2 and EDR3 parallaxes show that these measurements can change considerably between different observations \citep{Gaia18,Gaia20}. 

Our final mass census contains 750 stars, either with a mass estimate or a spectral type, which we use to derive masses. NGC\,2264 has an upper age estimate of around 5\,Myr, which means that most of the stars in this cluster are still pre-main sequence stars. 
To convert pre-main sequence spectral types into mass estimates, we use the values in Table~2 from \citet{Kirk10} who follow the procedure introduced in \citet{Luhman03b} to estimate masses based on effective temperature.

However, the surveys we use to construct the mass census contain more stars identified as members but that do not have masses or spectral types and with membership probabilities below 50 per cent. The size of our mass census should therefore not be considered as equal to the current size of the cluster population. Depending on the level of interstellar extinction and how embedded the sources are, current observations are also unlikely to probe the full mass range down to 0.1\,M$_\odot$, where a large number of possible member stars might still remain undetected. \citet{Venuti19} identified a total of 1369 sources in the clustered population, of which only 54 stars had a clear identification as field stars. The others were either ``very likely'' or ``possible'' members of the cluster or had no membership flag at all. \citet{Teixeira12} estimated the size of the stellar population in NGC\,2264 within their search fields that contain all three of the regions of interest to contain 1436$\pm$242 members. This highlights that the current size of NGC\,2264 is likely in the range of 1200 -- 1450 members.

\begin{figure*}
\begin{center}
\rotatebox{270}{\includegraphics[scale=0.75]{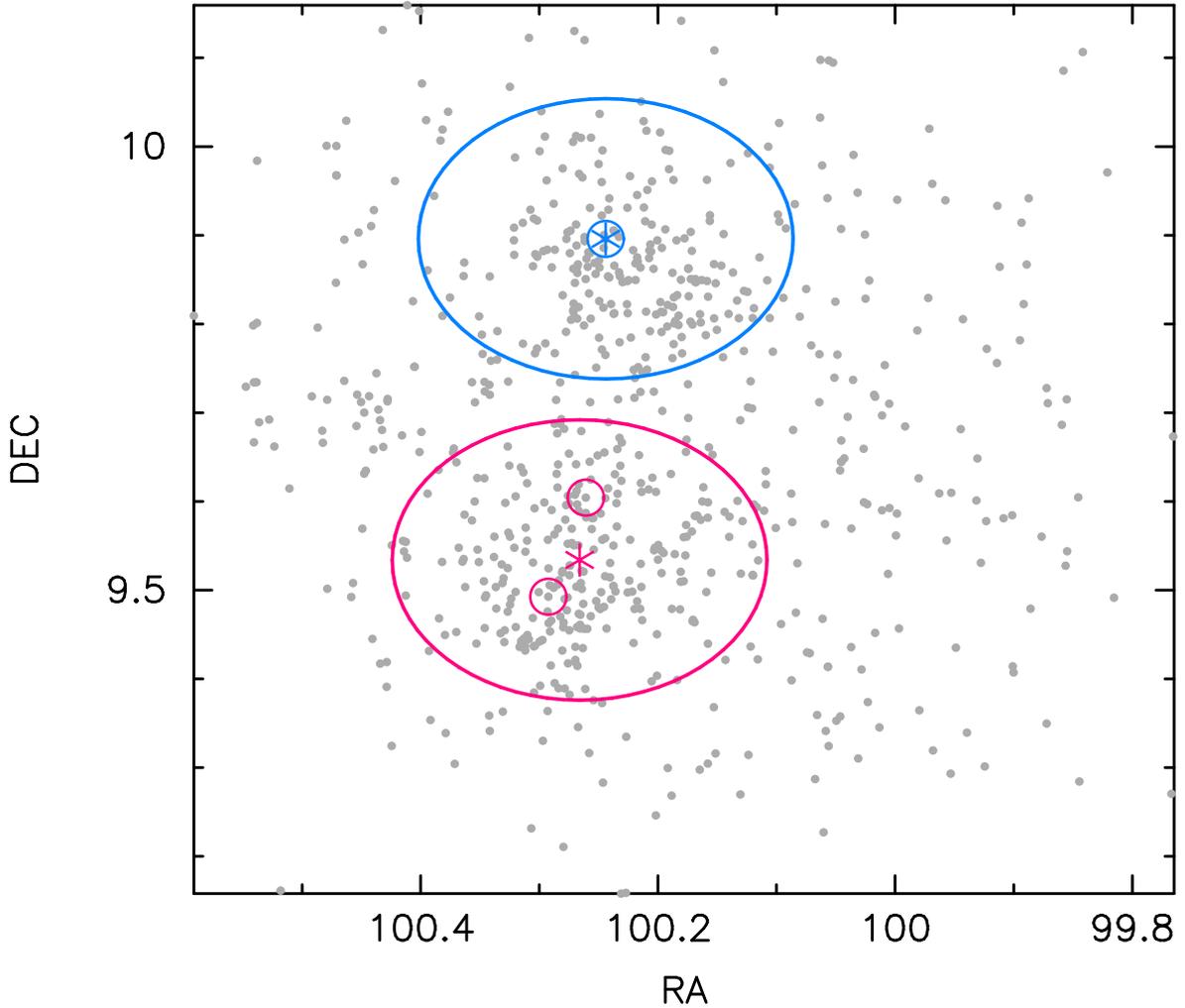}}
\end{center}
\caption[bf]{Map of NGC\,2264 with the stars in our sample shown by the grey points. The most massive star in the upper subcluster (S~Mon) is shown by the smaller blue circle, and the extent over which we perform the structural analysis for this subcluster is indicated by the larger blue circle (which has a radius of 0.158 degrees, or $\sim$2\,pc). In the lower subcluster, the massive stars IRS~1 and IRS~2 are indicated by the small pink circles, and the region over which we perform the structural analysis is shown by the larger pink circle (which has a radius of 0.158 degrees, or $\sim$2\,pc). The central position of this circle is indicated with an asterisk.}
\label{Full_map}
\end{figure*}

We show a map of the data in Fig.~\ref{Full_map}. The 750 stars in our sample are indicated by the grey points. We split the region into two sub-clusters,  based on the observed age dichotomy in NGC\,2264. In the upper, northern cluster surrounding the massive star S~Mon, the stars have ages around 5\,Myr, and in the lower, southern cluster surrounding the stars IRS~1 and IRS~2, the stars have ages around 2\,Myr. Our assumption is that -- whilst the stars may have formed from the same molecular cloud -- there have been at least two epochs of star formation in NGC\,2264 and we therefore split the region into two subclusters in order to perform our analysis.

In Fig.~\ref{Full_map} we show the position of S~Mon by the small blue circle, and the extent over which we perform our structural analyses by the larger blue circle, which has a radius of 0.168 degrees ($\sim$2\,pc at the adopted distance to NGC\,2264). 
Similarly, we show the positions of IRS~1 and IRS~2 in the southern subcluster by the small pink circles, and the extent over which we perform the structural analyses by the larger pink circle, which again has a radius of 0.158 degrees (we indicate the centre of this larger circle by the pink asterisk).  

\section{Method}
\label{methods}

In this section we describe the methods used to quantify the structure in NGC\,2264 and in the $N$-body simulations, as well as the set-up of the $N$-body simulations. 

\subsection{Structural analyses}

\subsubsection{The $\mathcal{Q}$-parameter}

We quantify the overall structure of each subcluster using the $\mathcal{Q}$-parameter \citep{Cartwright04,Cartwright09b,Lomax11,Jaffa17}. The $\mathcal{Q}$-parameter is determined by constructing a minimum spanning tree (MST), which is a graph joining all of the points in a distribution via the shortest possible total path length and where there are no closed loops. We also construct the complete graph, in which every point is joined to every other point via a connecting line. 

The $\mathcal{Q}$-parameter is the mean length of the MST, $\bar{m}$, divided by the mean edge length of the complete graph, $\bar{s}$:
\begin{equation}
\mathcal{Q} = \frac{\bar{m}}{\bar{s}}.
\end{equation}
In two dimensions, $\mathcal{Q}<0.7$ indicates a clumpy, or substructured distribution, whereas $\mathcal{Q}>0.9$ indicates a smooth, centrally concentrated distribution. $\mathcal{Q}$ values in the range 0.7 -- 0.9 indicate a distribution with neither a substructured nor a centrally concentrated distribution. Such a distribution may be uniform, but in reality an observed distribution (or from simulations) may be more complex.  

The $\mathcal{Q}$-is formally scale-free, as $\bar{m}$ is normalised to
\begin{equation}
\frac{\sqrt{NA}}{N-1},
\end{equation}
where $N$ is the number of points in the distribution and $A$ is the area of a circle with a radius $R$ equal to the magnitude of the furthest point from the origin. $\bar{s}$ is normalised to the radius $R$.

The normalisation to the radius generally means that the $\mathcal{Q}$-parameter is scale-free. There is a mild dependence of $\mathcal{Q}$ on the number of stars \citep{Lomax11,Parker15c,Parker18a}, but the main degeneracy is when $\mathcal{Q}$ is calculated for a dataset with significant outliers, e.g.\,\, stars that are tens of pc away from the main part of a star-forming region with a size scale of $\sim$1\,pc. This is potentially an issue for our comparison with $N$-body simulations \citep{Parker14e} and so we limit our analysis in our simulations to stars within two half-mass radii of the centre of the region.  

\subsubsection{The $\Lambda_{\rm MSR}$ mass segregation ratio}

We quantify the degree of mass segregation using the \citet{Allison09a} $\Lambda_{\rm MSR}$ ratio. This constructs an MST between $N_{\rm MST}$ stars in a chosen subset to determine the  length $l_{\rm subset}$ of this MST.  We then compare this length with the average MST length of $N_{\rm MST}$ randomly chosen stars, $\langle l_{\rm average}\rangle$:
\begin{equation}
\Lambda_{\rm MSR} = \frac{\langle l_{\rm average}\rangle}{l_{\rm subset}}^{+\sigma_{5/6}/{l_{\rm subset}}}_{-\sigma_{1/6}/{l_{\rm subset}}},
\end{equation}
where the lower (upper) uncertainty is the MST length which lies 1/6 (5/6) of the way through an ordered list of all the random lengths, which corresponds to a 66 per cent deviation from the median value, $\langle l_{\rm average}\rangle$ \citep{Allison09a,Parker11b,Parker18a}. This has the advantage that an outlying datapoint cannot strongly influence the uncertainty, although we in any case restrict our determination of $\Lambda_{\rm MSR}$ in the simulations to stars within two half-mass radii of the centre. 

\subsubsection{Local surface density}

The local surface density, $\Sigma$, is determined for every star using the equation in \citet{Casertano85}:
\begin{equation}
\Sigma = \frac{N - 1}{\pi r_N^2},
\end{equation}
where $N$ is the number of nearest neighbours and $r_N$ is the projected distance to the $N^{\rm th}$ nearest neighbour. The value of $\Sigma$ is fairly insensitive to the choice of $N$ \citep{Bressert10,Parker12d}, and we adopt $N = 10$ here. The advantage of $\Sigma$ is that it can be used to directly compare the densities in simulated star-forming regions with observed regions. 

\citet{Maschberger11} introduced the $\Sigma - m$ plot to quantify differences between the surface density of the most massive stars and the surface density of all of the stars in a region. \citet{Kupper11} and \citet{Parker14b} introduced the local density ratio, $\Sigma_{\rm LDR}$ as the ratio between the median surface density of the ten most massive stars, $\tilde{\Sigma}_{\rm 10}$,  (or any other defined subset) and the median surface density of all of the stars in the region, $\tilde{\Sigma}_{\rm all}$:
\begin{equation}
\Sigma_{\rm LDR} = \frac{\tilde{\Sigma}_{\rm 10}}{\tilde{\Sigma}_{\rm all}},
\end{equation}
the significance of which is quantified using a Kolmogorov-Smirnov test.

\subsubsection{Binary stars}

Few structural analyses using the $\mathcal{Q}$-parameter, $\Lambda_{\rm MSR}$ and $\Sigma_{\rm LDR}$ have been performed on distributions with a significant fraction of binary systems \citep[although this is common in studies that use the 2-point correlation function, e.g.][]{Larson95,Simon97,Kraus08}. \citet{Kupper11} showed that a population of binary systems will slightly alter the $\mathcal{Q}$-parameter, by virtue of the extra (short) links present between stars. This is more complicated for observational data where information on the binary systems may be restricted to binaries with a certain range of separations \citep[e.g.][]{King12a,King12b,Duchene13b}. In Appendix~\ref{appendix:binaries} we show that the range of binary separations we are sensitive to has a negligible effect on the results.
 
\subsection{$N$-body simulations}
\label{nbody_method}

We set up and run $N$-body simulations of star-forming regions designed to mimic the subclusters in NGC\,2264. In the observational sample, the two subclusters contain 212 and 258 objects within our chosen sub-cluster boundaries. However, our $N$-body simulations have a higher number of systems (725 systems for each of the two subclusters), for several reasons. 

First, the observational sample we use for our analysis is limited to stars for which we have a mass estimate or could derive one from their spectral type. Literature estimates considering all possible members for NGC\,2264 give suggest considerably higher numbers, which are within the numbers chosen for the $N$-body simulations \citep{Teixeira12,Venuti19}. Second, the observational samples are likely missing a proportion of stars with very low masses (0.1--0.3 \,M$_\odot$) due to extinction or the sources being embedded. Third, we only use observations from the mass census that are within a  2 pc radius around the centre of each subcluster, and so there will be stars that originated from these subclusters that have travelled beyond this (arbitrary) boundary, but are still bound to the subclusters. Typically, our $N$-body simulation expand and there are around 200 - 300 stars within the 2\,pc radius.

\subsubsection{Stellar systems}

Our simulations each have {\bf $N = 725$} stellar systems, with masses drawn from the \citet{Maschberger13} Initial Mass Function (IMF) with a probability distribution function of the form
\begin{equation}
p(m) \propto \left(\frac{m}{\mu}\right)^{-\alpha}\left(1 + \left(\frac{m}{\mu}\right)^{1 - \alpha}\right)^{-\beta},
\label{maschberger_imf}
\end{equation}
where $\mu = 0.2$\,M$_\odot$ is the average stellar mass, $\alpha = 2.3$ and $\beta = 1.4$. We sample this IMF between 0.1 and  50\,M$_\odot$.

\begin{table*}
\caption[bf]{Binary properties of systems in our simulations, which are set up to mimic the distributions observed in the Galactic field. We show the spectral type of the primary mass, the main sequence mass range this corresponds to, the binary fraction $f_{\rm bin}$, and for stars less massive than 3\,M$_\odot$ we show the mean separation $\bar{a}$, and the mean (${\rm log}\,\bar{a}$) and variance ($\sigma_{{\rm log}\,\bar{a}}$) of the log-normal fits to these distributions. OB stars ($m_p>$3.0\,M$_\odot$) are not drawn from a log-normal distribution but instead are drawn from a log-uniform \citet{Opik24} distribution in the range 0 - 50\,au.}
\begin{center}
\begin{tabular}{|c|c|c|c|c|c|c|}
\hline 
Spectral Type & Primary mass & $f_{\rm bin}$ & $\bar{a}$ & ${\rm log}\,\bar{a}$ & $\sigma_{{\rm log}\,\bar{a}}$ & Ref. \\
\hline
M-dwarf & $0.10 < m_p/$M$_\odot \leq 0.45$ & 0.34 & 16\,au & 1.20 & 0.80 & \citet{Bergfors10,Janson12} \\
\hline 
F, G, K & $0.45 < m_p/$M$_\odot \leq 1.20$ & 0.46 & 50\,au & 1.70 & 1.68 & \citet{Raghavan10} \\
\hline
A & $1.20 < m_p/$M$_\odot \leq 3.00$ & 0.48 & 389\,au & 2.59 & 0.79 & \citet{DeRosa14} \\
\hline
OB & $m_p>$ 3.00\,M$_\odot$ & 1.00 & {\"O}pik & 0 -- 50\,au & log-uniform & \citet{Sana13} \\ 
\hline
\end{tabular}
\end{center}
\label{bin_props}
\end{table*}

We then select a random number $\mathcal{R}_n$ between 0 and 1 to determine whether a system is a single or binary. The binary fraction, $f_{\rm bin}$ is defined as 
\begin{equation}
f_{\rm bin} = \frac{B + \ldots}{S + B + \ldots},
\end{equation} 
where $S$ and $B$ are the numbers of single and binary systems, respectively \citep[we do not include triple or quadruple systems in the simulations, although they are certainly an important outcome of the star formation process,][]{Tokovinin08,Tokovinin18,Reipurth14,Pineda15}. 

In accordance with observations in the Galactic field we assign a binary fraction as a function of the stellar mass. For stars $m > 3.00$\,M$_\odot$, $f_{\rm bin} = 1.00$ \citep{Mason98,Kouwenhoven07,Sana13}, for stars $1.20 \leq m \leq 3.00$, $f_{\rm bin} = 0.48$ \citep{DeRosa14}, for stars $0.84 \leq m < 1.20$, $f_{\rm bin} = 0.46$ \citep{Raghavan10}, for stars $0.45 \leq m < 0.84$, $f_{\rm bin} = 0.45$ \citep{Mayor92} and for stars   $0.10 \leq m < 0.45$, $f_{\rm bin} = 0.34$ \citep{Janson12}.  If  $\mathcal{R}_n <  f_{\rm bin}$ for the mass range the star falls within then the stellar system is set to be a binary. 

The distribution of binary semimajor axes is also a function of primary mass. Binaries with $m > 3.00$\,M$_\odot$ have semimajor axes drawn from a log-uniform \citet{Opik24} distribution between 0 and 50\,au, whereas all other systems are drawn from a lognormal distribution where the mean and variance is also a function of the primary mass in the binary. These values are  summarised in Table~\ref{bin_props} but are characterised by higher mean semimajor axes for more massive primary stars, moving to progressively lower mean semimajor axes for low-mass stars \citep{DeRosa14,Raghavan10,Ward-Duong15,Janson12}.

We determine the mass of the companion star in the binary by assuming a flat mass ratio distribution as observed in the Galactic field \citep{Reggiani11a,Reggiani13}. Eccentricities are also drawn from a flat distribution \citep{Raghavan10,Duchene13b}, apart from binaries with semimajor axes less than 1\,au, which are placed on circular ($e = 0$) orbits.

\subsubsection{Star forming regions}

We distribute our stellar systems randomly in a box fractal distribution according to the method in \citet{Goodwin04a}. This method is detailed in many other works \citep[e.g.][]{Allison10,Parker14b,Schoettler20} and we refer the interested reader to those papers for details. In short, the amount of spatial substructure in the region is set by the fractal dimension, $D$, with lower values  (e.g. $D = 1.6$) corresponding to more substructure, whereas $D = 3.0$ is a uniform spherical distribution. 

The fractals are constructed by determining how many parent particles mature and produce child particles, with the substructured regions producing fewer child particles.  Stellar velocities for the parent particles are drawn from a Gaussian with mean zero, and the children inherit these velocities plus a small random noise component that decreases with each successive generation of children in the fractal. This way, the stars in regions with a high degree of substructure have very correlated velocities on local scales, but on large scales the velocities can vary significantly, similar to the observed \citet{Larson81} relations. 

\begin{table}
  \caption[bf]{Summary of the initial conditions adopted for our simulations. The total number of stellar systems (single and binary) is fixed at $N = 750$ but we vary the initial radius $r_F$, fractal dimension $D$ and virial ratio $\alpha_{\rm vir}$. The different combinations of radii and fractal dimension give different initial median stellar densities $\tilde{\rho}$, which in turn result in different (observable) median stellar surface densities, $\tilde{\Sigma}$. }
  \begin{center}
    \begin{tabular}{|c|c|c|c|c|c|}
      \hline
Sim. ID & $r_F$ & $D$ & $\alpha_{\rm vir}$ & $\tilde{\rho}$ (0\,Myr) & $\tilde{\Sigma}$ (0\,Myr) \\
\hline
16-03-1 & 1\,pc & $1.6$ & 0.3 & 10\,000\,M$_\odot$\,pc$^{-3}$ & 3000\,stars\,pc$^{-2}$ \\
16-05-1 & 1\,pc & $1.6$ & 0.5 & 10\,000\,M$_\odot$\,pc$^{-3}$ & 3000\,stars\,pc$^{-2}$ \\
30-03-1 & 1\,pc & $3.0$ & 0.3 & 150\,M$_\odot$\,pc$^{-3}$ & 400\,stars\,pc$^{-2}$ \\
30-05-1 & 1\,pc & $3.0$ & 0.5 & 150\,M$_\odot$\,pc$^{-3}$ & 400\,stars\,pc$^{-2}$ \\
16-03-5 & 5\,pc & $1.6$ & 0.3 & 70\,M$_\odot$\,pc$^{-3}$ & 100\,stars\,pc$^{-2}$ \\
16-05-5 & 5\,pc & $1.6$ & 0.5 & 70\,M$_\odot$\,pc$^{-3}$ & 100\,stars\,pc$^{-2}$ \\
20-03-5 & 5\,pc & $2.0$ & 0.3 & 10\,M$_\odot$\,pc$^{-3}$ & 40\,stars\,pc$^{-2}$ \\
20-05-5 & 5\,pc & $2.0$ & 0.5 & 10\,M$_\odot$\,pc$^{-3}$ & 40\,stars\,pc$^{-2}$ \\
      \hline
    \end{tabular}
  \end{center}
  \label{simulations}
\end{table}

Finally, the velocities of individual stars are scaled to the overall `bulk motion' of the star-forming regions, as set by the virial ratio $\alpha_{\rm vir} = |\Omega|/\mathcal{K}$, where $\Omega$ and $\mathcal{K}$ are the total potential and kinetic energies, respectively. In accordance with observations that suggest star-forming regions are initially subvirial \citep[e.g.][]{Peretto06,Foster15}, we set some simulations with $\alpha_{\rm vir} = 0.3$, whereas others are set to $\alpha_{\rm vir} = 0.5$ (virial equilibrium).  A summary of the initial conditions is given in Table~\ref{simulations}.

We evolve the star-forming regions using the \texttt{kira} integrator in \texttt{Starlab} \citep{Zwart99,Zwart01}, which is a 4$^{\rm th}$-order Hermite $N$-body code with block time-stepping. We include stellar and binary evolution in the simulations by utilising \texttt{SeBa} look-up tables, also in \texttt{Starlab} \citep{Zwart96,Zwart12}. The simulations are evolved for 10\,Myr, i.e.\,\,well beyond the upper age estimate for the S~Mon subcluster (5\,Myr). 

\section{Results}
\label{results}

\subsection{Observational data}

\subsubsection{S~Mon}

We first determine the $\mathcal{Q}$-parameter and its constituent $\bar{m}$ and $\bar{s}$ for the S~Mon subcluster.  $\mathcal{Q} = 0.83$ and $\bar{m} = 0.51$ and $\bar{s} = 0.61$, which is at the boundary between a substructured and a smooth distribution, although the plot of $\bar{m} - \bar{s}$ \citep{Cartwright09b,Parker18a} tentatively indicates a fractal distribution with $D = 2.6$.   

In Fig.~\ref{obs_lambda_MSR-a} we show the mass segregation ratio $\Lambda_{\rm MSR}$ as a function of the $N_{\rm MST}$ most massive stars. This plot clearly shows no significant deviation from unity, meaning that the most massive stars have the same spatial distribution as the average stars. 

We show the local surface density as a function of mass for each star in Fig.~\ref{obs_sigma_LDR-a}. The median surface density for the region is 21\,stars\,pc$^{-2}$ and the most massive stars have very similar values (compare the solid red line with the blue dashed line). 

\subsubsection{IRS\,1/2}

We now focus on the subcluster centred between the stars IRS\,1 and IRS\,2. The $\mathcal{Q}$-parameter and associated $\bar{m}$ and $\bar{s}$ values are 0.80, 0.61 and  0.76, respectively. Whilst this $\mathcal{Q}$-parameter is very much in the no-man's land between being categorized as either substructured or smooth, the $\bar{m}$ and $\bar{s}$ values place it in the smooth, centrally concentred regime on the \citet{Cartwright09b} plot. 

\begin{figure*}
  \begin{center}
\setlength{\subfigcapskip}{10pt}
\hspace*{0.3cm}\subfigure[S~Mon]{\label{obs_lambda_MSR-a}\rotatebox{270}{\includegraphics[scale=0.3]{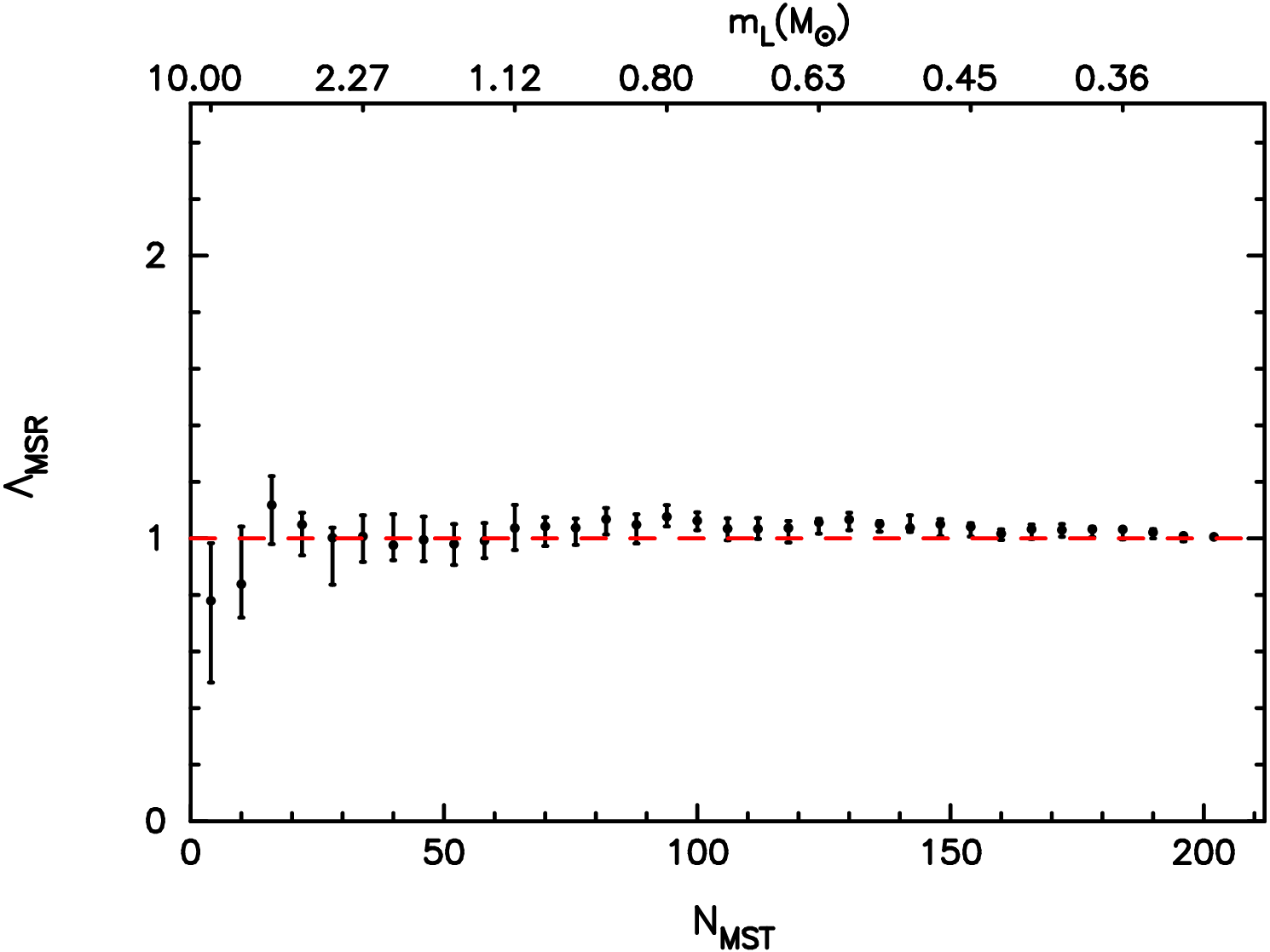}}}
\hspace*{0.3cm}\subfigure[IRS\,1/2]{\label{obs_lambda_MSR-b}\rotatebox{270}{\includegraphics[scale=0.3]{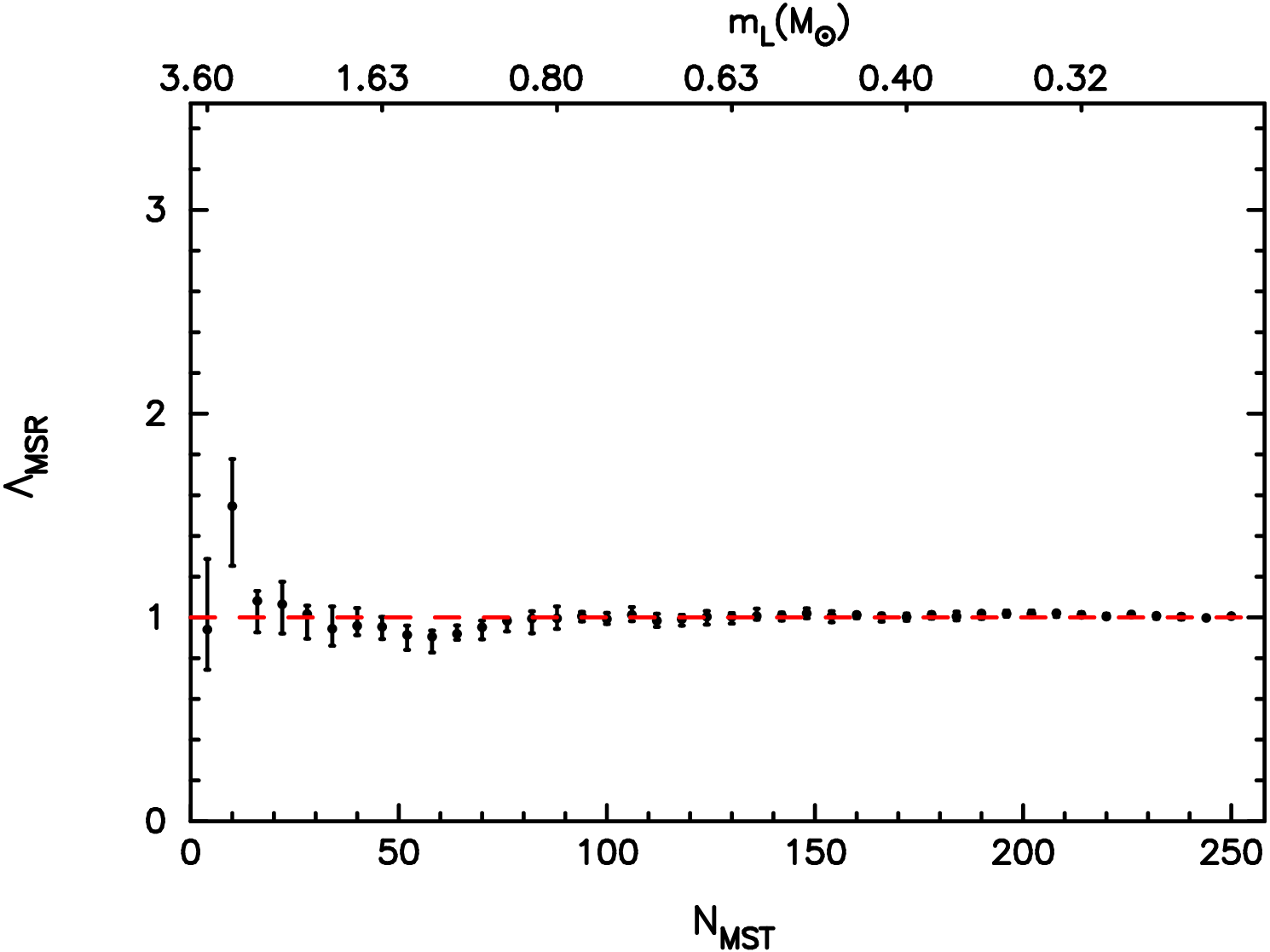}}}
\caption[bf]{The mass segregation ratio, $\Lambda_{\rm MSR}$, as a function of the $N_{\rm MST}$ most massive stars in each of the subclusters in NGC\,2264. The lowest mass star in the $N_{\rm MST}$ subset is indicated on the top axis. The horizontal red dashed line indicates $\Lambda_{\rm MSR} = 1$, i.e.\,\,no mass segregation.}  
\label{obs_lambda_MSR}
  \end{center}
\end{figure*}

In Fig.~\ref{obs_lambda_MSR-b} we show the mass segregation ratio $\Lambda_{\rm MSR}$ as a function of the $N_{\rm MST}$ most massive stars. The 4 most massive stars do not deviate from a mass segregation ratio of unity, and although the 10 most massive stars have $\Lambda_{\rm MSR} = 1.59^{+1.76}_{-1.24}$, we would caution against interpreting this as a significant signature of mass segregation \citep[see][]{Parker15b}.

\begin{figure*}
  \begin{center}
\setlength{\subfigcapskip}{10pt}
\hspace*{0.3cm}\subfigure[S~Mon]{\label{obs_sigma_LDR-a}\rotatebox{270}{\includegraphics[scale=0.35]{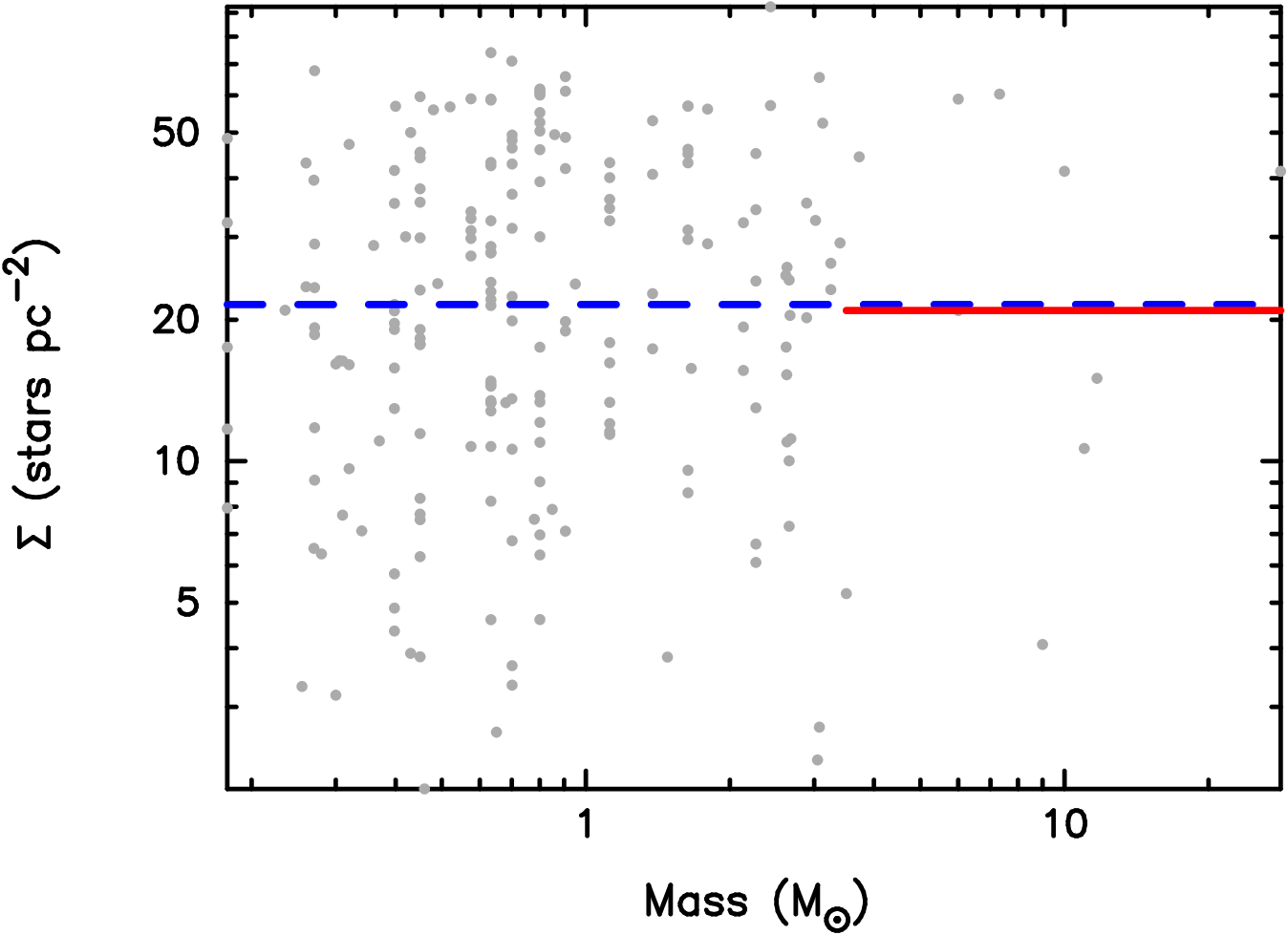}}}
\hspace*{0.3cm}\subfigure[IRS\,1/2]{\label{obs_sigma_LDR-b}\rotatebox{270}{\includegraphics[scale=0.35]{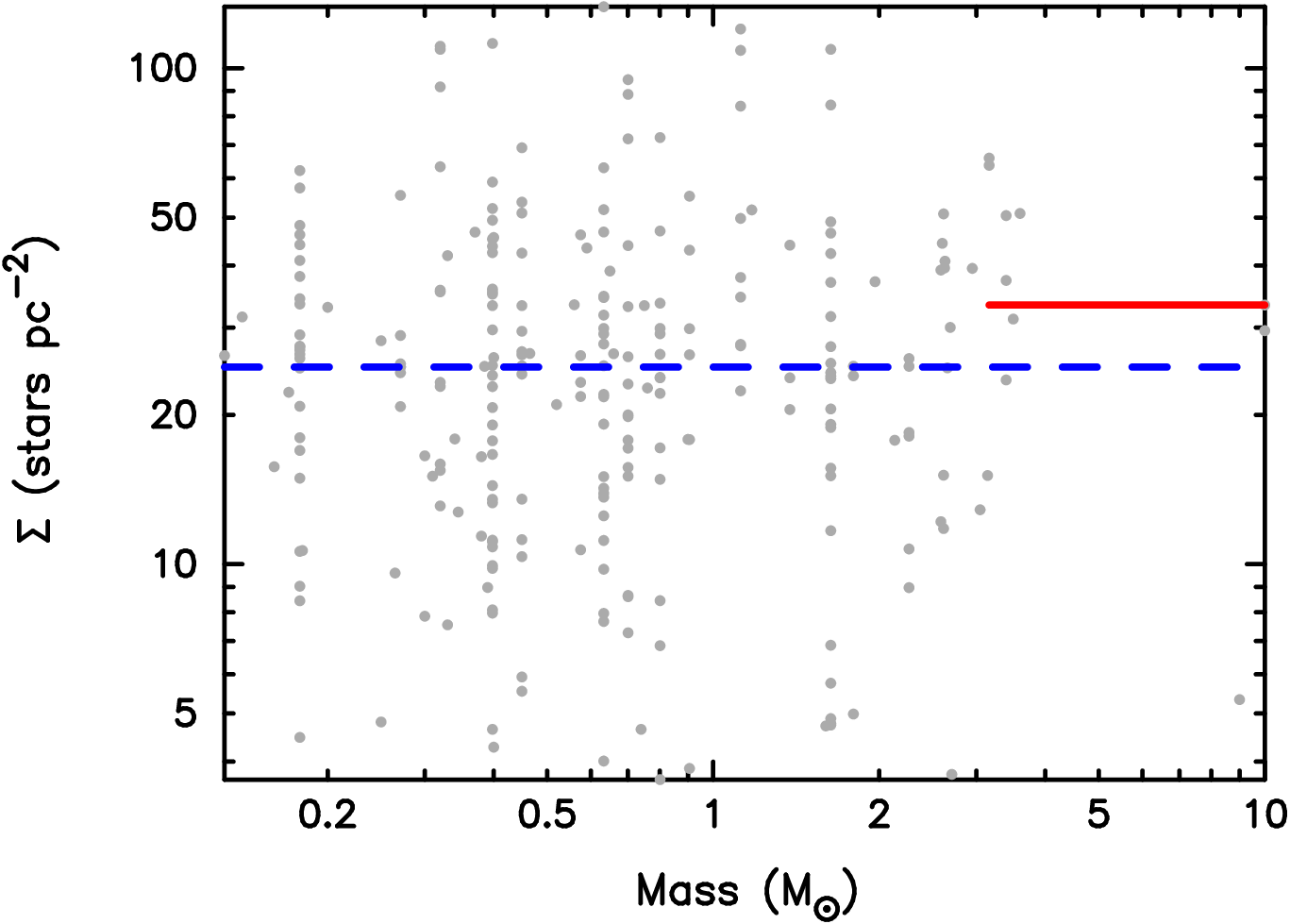}}}
\caption[bf]{The local surface density, $\Sigma$, as a function of stellar mass in each of the subclusters in NGC\,2264. The blue dashed line is the median surface density for all of the stars in the subcluster, and the solid red line is the median surface density for the ten most massive stars. The ratio of these  median densities is used to calculate the local surface density ratio, $\Sigma_{\rm LDR}$. }
\label{obs_sigma_LDR}
  \end{center}
\end{figure*}

The median surface density for the IRS\,1/2 subcluster is $\tilde{\Sigma} = 24$\,stars\,pc$^{-2}$, similar to the S~Mon subcluster. However, the 10 most massive stars in IRS\,1/2 do display a significant difference in their surface densities, as shown in Fig.~\ref{obs_sigma_LDR-b} (compare the median surface density of the massive stars, shown by the red line, to the median surface density for all stars, shown by the blue dashed line). The KS test returns a $p$-value $p= 3.6 \times 10^{-2}$ that the two samples share the same underlying parent distribution.

\begin{table*}
\caption[bf]{Summary of the results for the observed subclusters in NGC\,2264. We show the number of stars in our dataset for each subcluster, $N$, the age, the mean length of the minimum spanning tree of each region, $\bar{m}$, the mean length of the complete graph for each region, $\bar{s}$, the $\mathcal{Q}$-parameter, the mass segregation ratio for the four most massive stars, $\Lambda_{\rm MSR, 4}$, the mass segregation ratio for the ten most massive stars, $\Lambda_{\rm MSR, 10}$, the median surface density for all stars in each subcluster, $\tilde{\Sigma}$, the local surface density ratio for the ten most massive stars,  $\Sigma_{\rm LDR, 10}$, and the KS test results between the surface densities of the ten most massive stars and the rest. }
\begin{center}
\begin{tabular}{|c|c|c|c|c|c|c|c|c|c|c|c|}
\hline 
Subcluster & $N$ & Age & $\bar{m}$ & $\bar{s}$ & $\mathcal{Q}$ & $\Lambda_{\rm MSR, 4}$ & $\Lambda_{\rm MSR, 10}$ & $\tilde{\Sigma}$ & $\Sigma_{\rm LDR, 10}$ & KS test \\
\hline
S~Mon & 212 & 5\,Myr & 0.51 & 0.61 & 0.83 & 0.71$^{+0.80}_{-0.50}$ & 0.83$^{+0.98}_{-0.83}$ & 21\,stars\,pc$^{-2}$ & 0.97 & $D = 0.24$, $p = 0.56$\\
\hline
IRS~1/2 & 258 & 2\,Myr & 0.61 & 0.76 & 0.80 & 0.92$^{+1.45}_{-0.73}$ & 1.59$^{+1.76}_{-1.24}$ & 24\,stars\,pc$^{-2}$ & 1.33 & $D = 0.44$, $p = 3.6 \times 10^{-2}$\\
\hline
\end{tabular}
\end{center}
\label{obs_values}
\end{table*}

\subsection{$N$-body simulations}

\begin{figure*}
  \begin{center}
\setlength{\subfigcapskip}{10pt}
\hspace*{0.1cm}\subfigure[$\Sigma$]{\label{16-03-1-a}\rotatebox{270}{\includegraphics[scale=0.34]{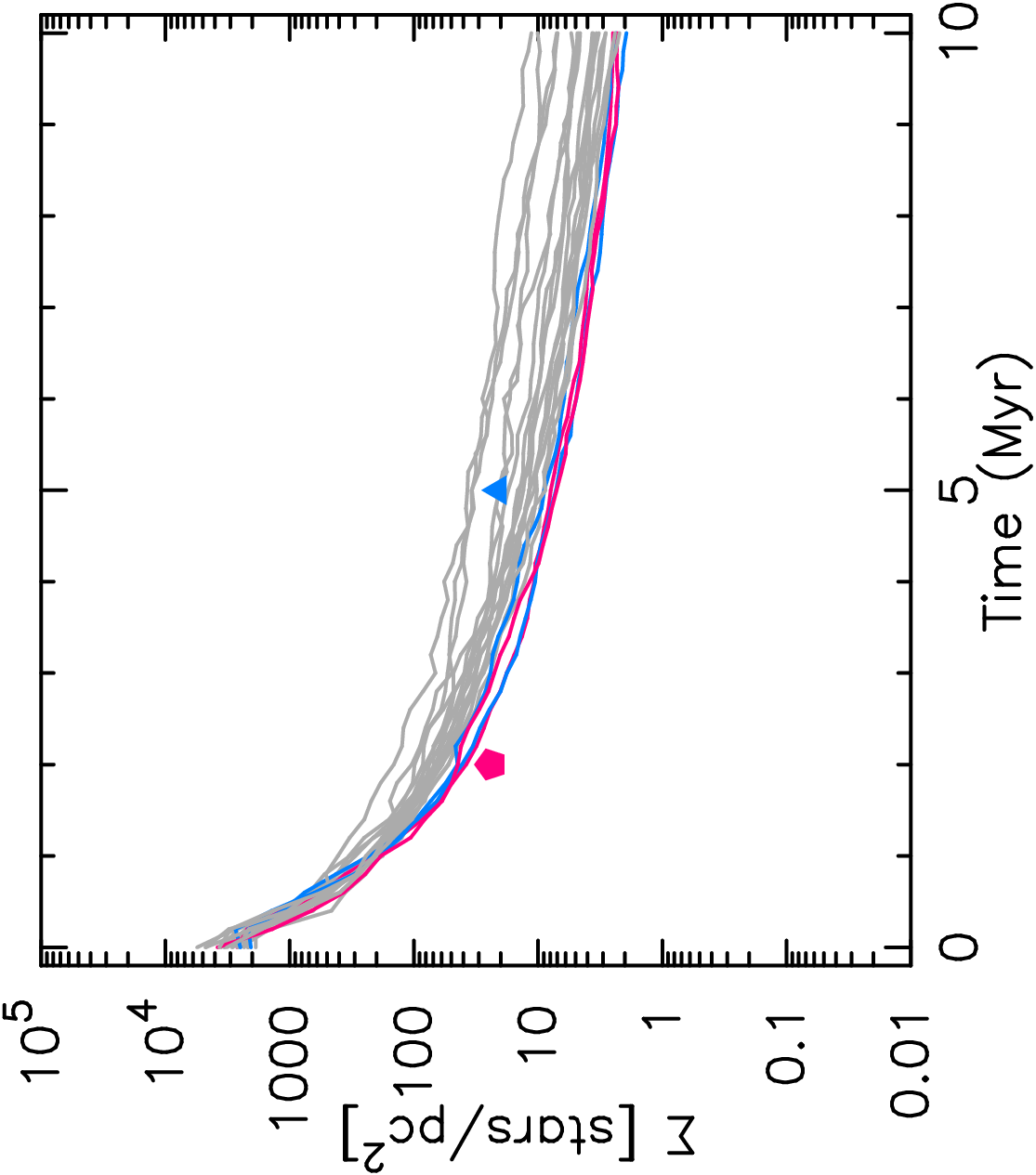}}} 
\hspace*{0.2cm}\subfigure[$\mathcal{Q}$-parameter]{\label{16-03-1-b}\rotatebox{270}{\includegraphics[scale=0.35]{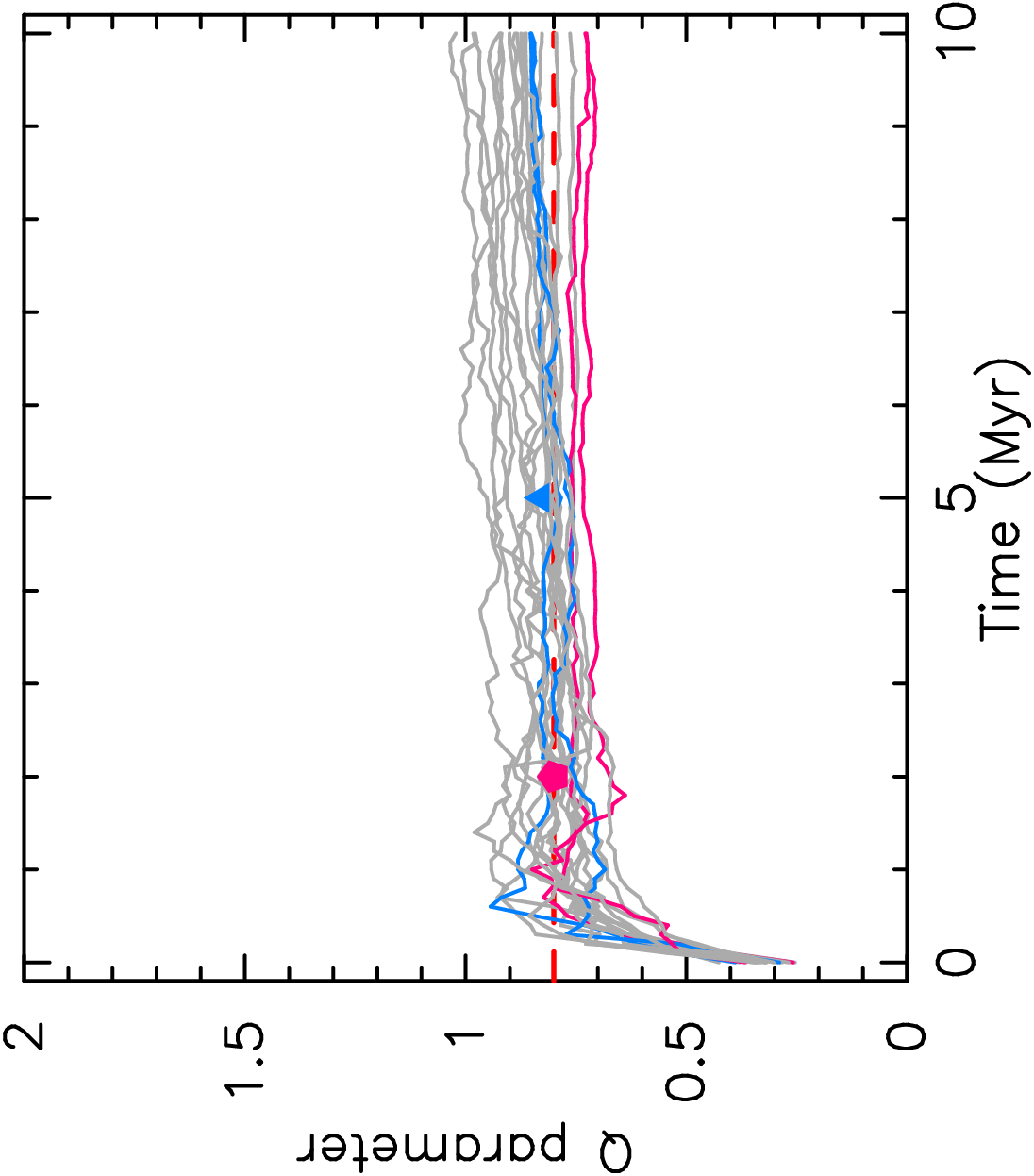}}}
\hspace*{0.1cm}\subfigure[$\mathcal{Q} - \Lambda_{\rm MSR}$]{\label{16-03-1-c}\rotatebox{270}{\includegraphics[scale=0.35]{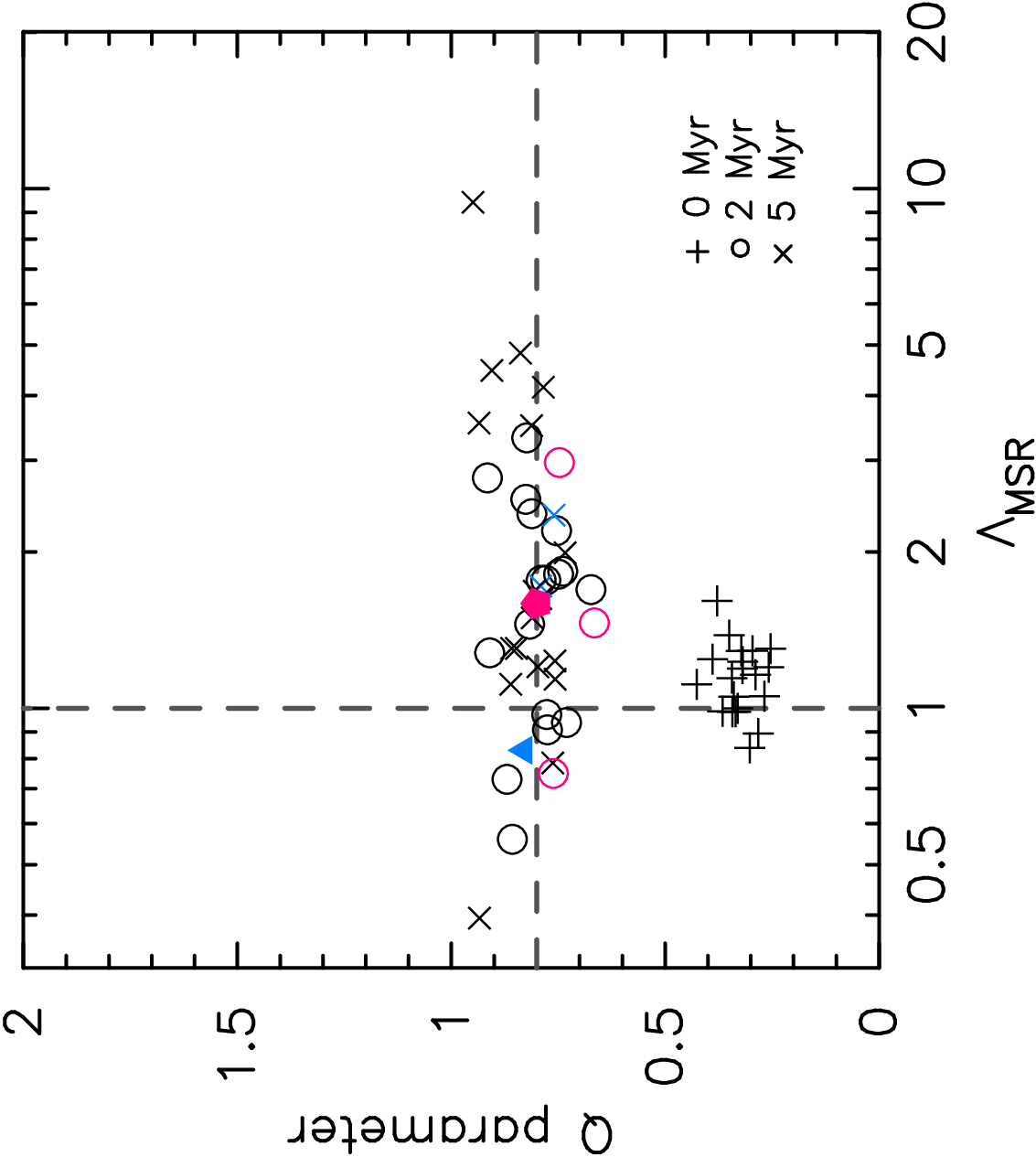}}}
\hspace*{0.1cm}\subfigure[$\mathcal{Q} - \Sigma_{\rm LDR}$]{\label{16-03-1-d}\rotatebox{270}{\includegraphics[scale=0.35]{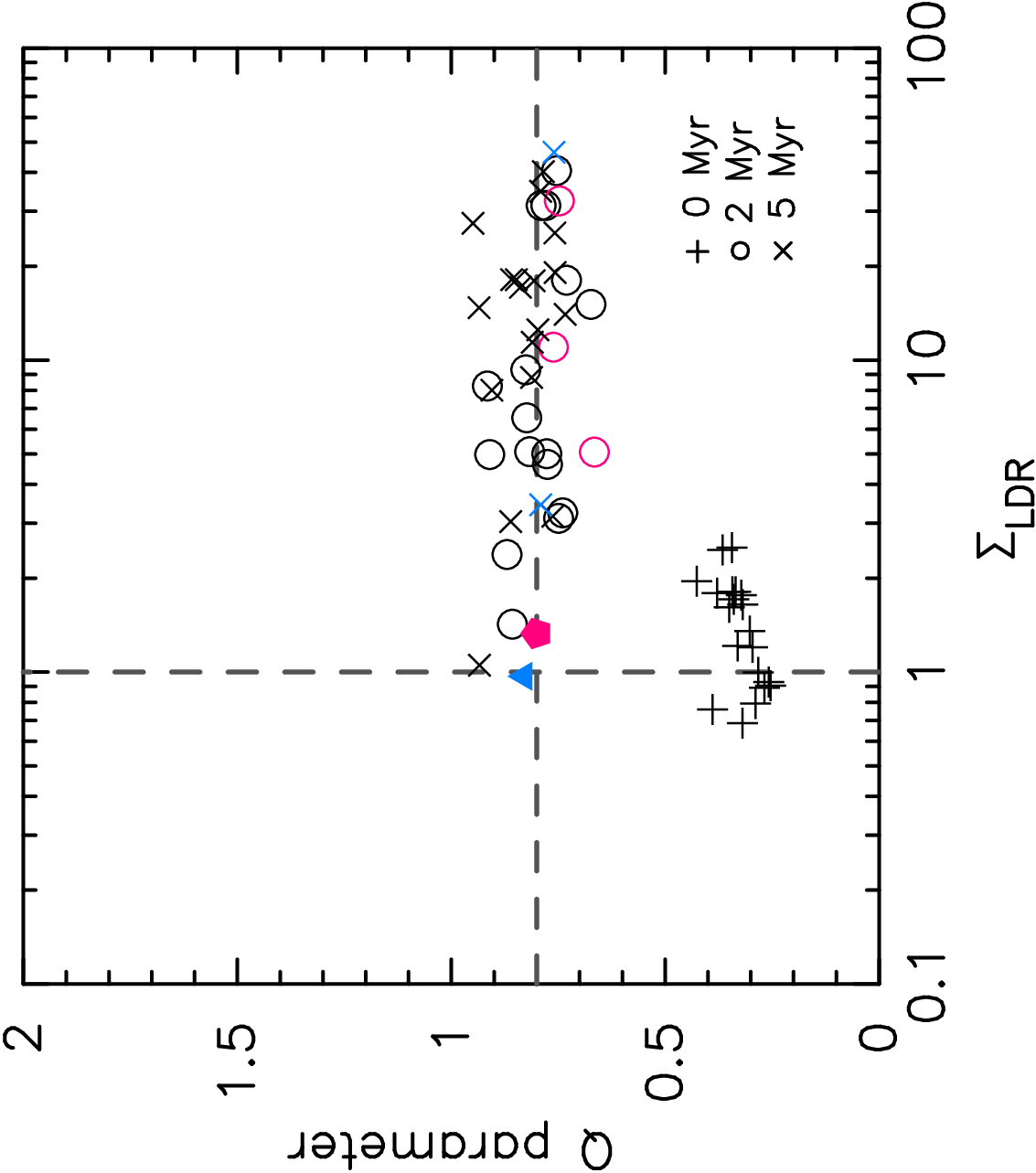}}}
\caption[bf]{Results from $N$-body simulations with highly substructured ($D = 1.6$), subvirial ($\alpha_{\rm vir} = 0.3$) initial conditions where the star-forming regions have initial radii of 1\,pc (labelled 16-03-1 in Table~\ref{simulations}). In all panels the pink pentagon corresponds to the observed values for the IRS\,1/2 subcluster and the blue triangle corresponds to the observed values for the S~Mon subcluster. The pink lines and symbols are simulations where the numbers of runaways and walkaways are consistent with the observed values for IRS\,1/2 at the upper age limit for this subcluster and the blue lines and symbols are simulations where the numbers of runaways and walkaways are consistent with the observed values for S~Mon at the upper age limit for this subcluster. Panel (a) shows the evolution of the median local surface density, $\tilde{\Sigma}$ and panel (b) shows the evolution of the $\mathcal{Q}$-parameter [the horizontal red dashed line indicates the boundary between substructured ($\mathcal{Q}<0.8$) and smooth ($\mathcal{Q}>0.8$) morphologies].  In panel (c) we show the $\mathcal{Q}$-parameter and $\Lambda_{\rm MSR}$ values at 0, 2 and 5\,Myr and in panel (d) we show the $\mathcal{Q}$-parameter and $\Sigma_{\rm LDR}$ values at 0, 2 and 5\,Myr. The horizontal lines in panels (c) and (d) show the boundary between substructured and smooth morphologies according to the $\mathcal{Q}$-parameter, and the vertical lines indicate $\Lambda_{\rm MSR} = 1$ (panel c) and $\Sigma_{\rm LDR} = 1$ (panel d).}
\label{16-03-1}
  \end{center}
\end{figure*}

We first provide an example of a simulation whose structural parameters ($\mathcal{Q}$-parameter, surface density $\Sigma$, mass segregation ratio $\Lambda_{\rm MSR}$ and local surface density ratio $\Sigma_{\rm LDR}$) provide a good fit to the observational data for the S~Mon and IRS~1/2 subclusters in NGC2264. These properties are shown in Fig.~\ref{16-03-1} for a simulation (`16-03-1' in Table~\ref{simulations}) with an initial radius of 1\,pc, a high degree of substructure ($D = 1.6$) and subvirial bulk motion ($\alpha_{\rm vir} = 0.3$). 

This simulation has a high initial surface density ($\Sigma = 3000$\,stars\,pc$^{-2}$), and undergoes subvirial collapse. As it collapses, the initial substructure is erased and the net effect is for the surface density to decrease (Fig~\ref{16-03-1-a}) and the $\mathcal{Q}$-parameter to increase (Fig.~\ref{16-03-1-b}). These simulations attain a small degree of mass segregation quantified as $\Lambda_{\rm MSR} > 1$, shown in Fig.~\ref{16-03-1-c} and the massive stars attain moderately high local surface densities ($\Sigma_{\rm LDR} > 1$, see Fig.~\ref{16-03-1-d}). 

In this figure, we show the observed values for S~Mon via the blue triangles, and for IRS\,1/2 via the pink heptagons. In general, the simulation data are shown by grey lines/points, save for when the individual simulations also produce the same numbers of runaway and walkaway stars to the observed subclusters \citep{Schoettler21b}. In that case, the simulations are shown by blue lines/points (S~Mon) or pink lines/points (IRS\,1/2). 

\begin{figure*}
  \begin{center}
\setlength{\subfigcapskip}{10pt}
\hspace*{0.1cm}\subfigure[$\Sigma$]{\label{20-03-5-a}\rotatebox{270}{\includegraphics[scale=0.34]{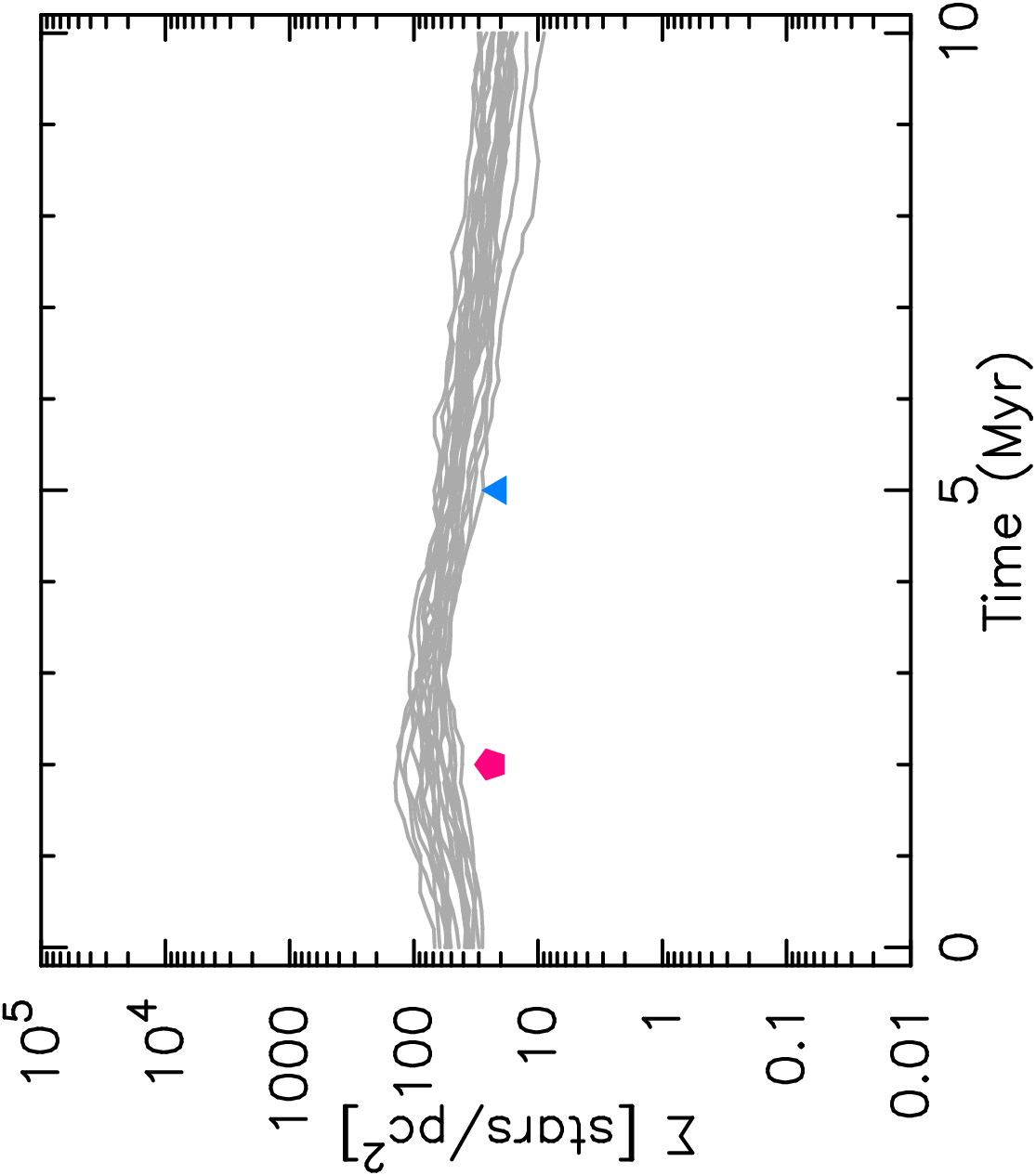}}} 
\hspace*{0.2cm}\subfigure[$\mathcal{Q}$-parameter]{\label{20-03-5-b}\rotatebox{270}{\includegraphics[scale=0.35]{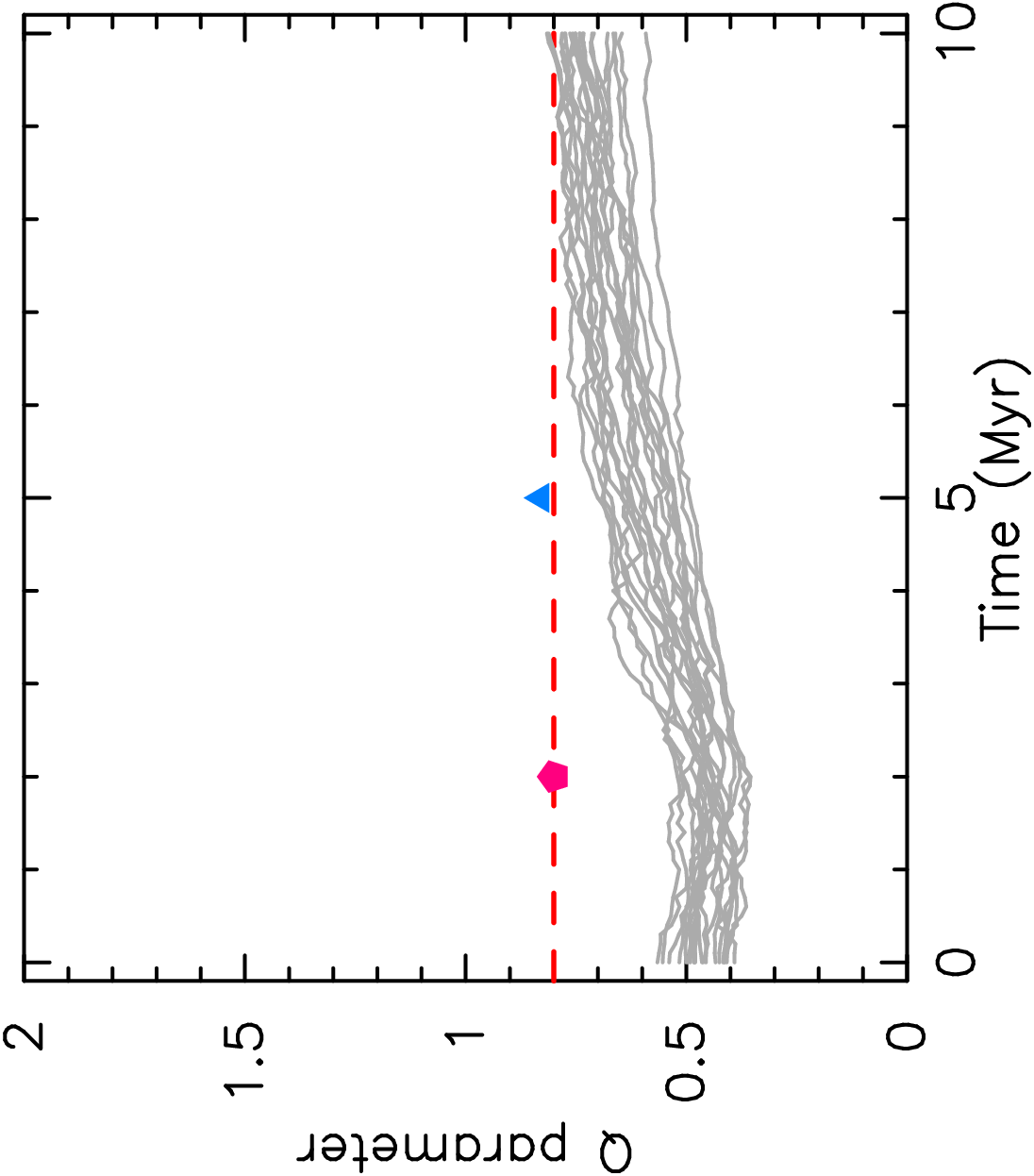}}}
\hspace*{0.1cm}\subfigure[$\mathcal{Q} - \Lambda_{\rm MSR}$]{\label{20-03-5-c}\rotatebox{270}{\includegraphics[scale=0.35]{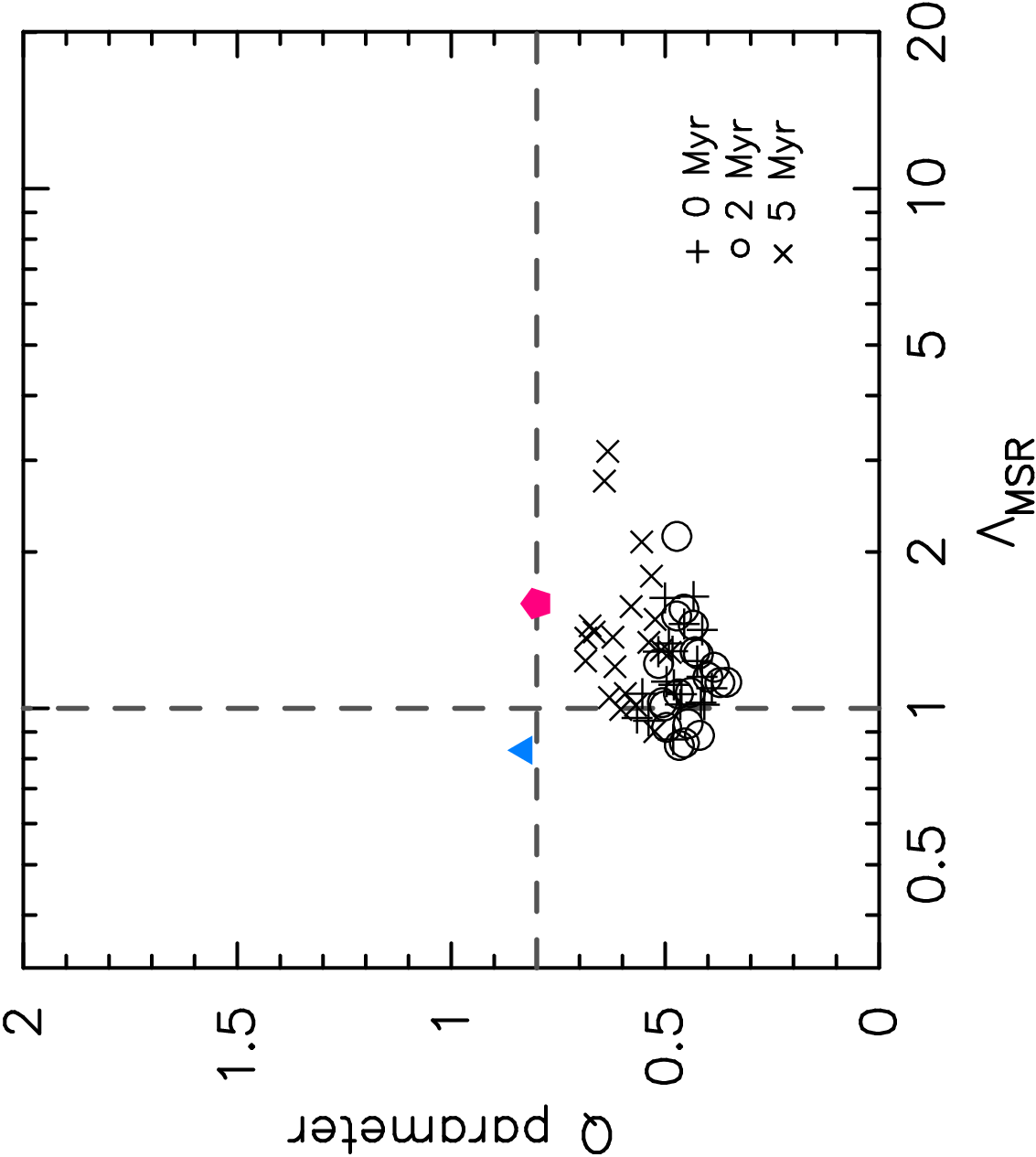}}}
\hspace*{0.1cm}\subfigure[$\mathcal{Q} - \Sigma_{\rm LDR}$]{\label{20-03-5-d}\rotatebox{270}{\includegraphics[scale=0.35]{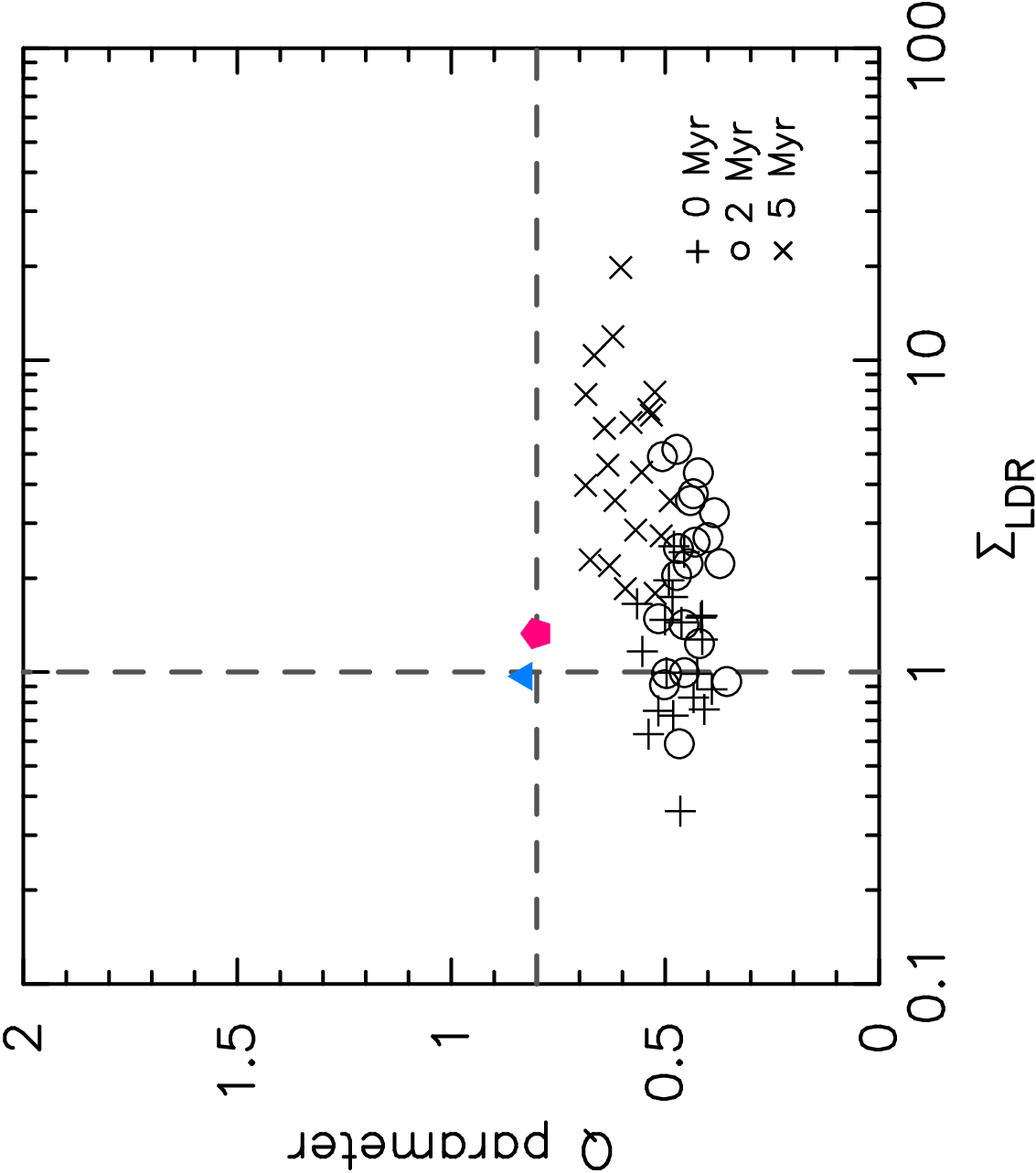}}}
\caption[bf]{Results from $N$-body simulations with moderately substructured ($D = 2.0$), subvirial ($\alpha_{\rm vir} = 0.3$) initial conditions where the star-forming regions have initial radii of 5\,pc (labelled 20-03-5 in Table~\ref{simulations}). In all panels the pink pentagon corresponds to the observed values for the IRS\,1/2 subcluster and the blue triangle corresponds to the observed values for the S~Mon subcluster. None of these simulations produce the observed numbers of runaways and walkways from these subclusters. Panel (a) shows the evolution of the median local surface density, $\tilde{\Sigma}$ and panel (b) shows the evolution of the $\mathcal{Q}$-parameter [the horizontal red dashed line indicates the boundary between substructured ($\mathcal{Q}<0.8$) and smooth ($\mathcal{Q}>0.8$) morphologies].  In panel (c) we show the $\mathcal{Q}$-parameter and $\Lambda_{\rm MSR}$ values at 0, 2 and 5\,Myr and in panel (d) we show the $\mathcal{Q}$-parameter and $\Sigma_{\rm LDR}$ values at 0, 2 and 5\,Myr. The horizontal lines in panels (c) and (d) show the boundary between substructured and smooth morphologies according to the $\mathcal{Q}$-parameter, and the vertical lines indicate $\Lambda_{\rm MSR} = 1$ (panel c) and $\Sigma_{\rm LDR} = 1$ (panel d).} 
\label{20-03-5}
  \end{center}
\end{figure*}

In Fig.~\ref{20-03-5} we show an example of a simulation in which the numbers of runaway and walkaway stars produced were categorically inconsistent with the numbers of observed RW/WW stars from both S~Mon and IRS~1/2. These simulations (labelled `20-03-5' in Table~\ref{simulations}) have an initial radius of 5\,pc, a moderate amount of spatial and kinematic substructure ($D = 2.0$) and subvirial bulk motion ($\alpha_{\rm vir} = 0.3$). 

The combination of large radius and moderate substructure means that these simulations take much longer to dynamically evolve, such that the substructure is still present after 10\,Myr of the simulation (note the low $\mathcal{Q}$-parameters in Fig.~\ref{20-03-5-b}, which are inconsistent with the observed values).  Whilst the median surface densities in these simulations are likely consistent with the observed value of S~Mon (the blue triangle in Fig.~\ref{20-03-5-a}), the observed $\mathcal{Q}$-parameters in the simulations strongly suggest that the two subclusters are more dynamically evolved. This also means that the simulations do not fall within the area occupied by the observations in the $\mathcal{Q} - \Lambda_{\rm MSR}$ (Fig.~\ref{20-03-5-c}) and $\mathcal{Q} - \Sigma_{\rm LDR}$ (Fig.~\ref{20-03-5-d}) plots. 

The two sets of simulations presented in Figs.~\ref{16-03-1}~and~\ref{20-03-5} are very much extrema within the full set of simulations. In Table~\ref{results-table} we provide an overview of whether other simulations are also consistent with both the numbers of runaways and walkaways (RW/WW), and the various structural diagnostics. For a parameter where the simulations are consistent with the observations, we place a `Y' in the table. For a parameter that is inconsistent with the observations, we place an `N', and where it is ambiguous we place an `?'. We then sum the number of positives (`Y's) to determine how consistent each simulation is with the observations and we deem the simulations with the maximum number of positives to be the best fit to the data. 

When we do this simplistic tallying, we find that the simulations most consistent with the observations are those that start with compact ($r_F = 1$\,pc) and substructured ($D = 1.6$) initial conditions, which are very dense ($10^4$\,M$_\odot$\,pc$^{-3}$).  The initial bulk motion can either be subvirial ($\alpha_{\rm vir} = 0.3$) or in virial equilibrium ($\alpha_{\rm vir} = 0.5$); in fact, this is a weak constraint because the local velocity dispersions in the substructure are usually highly subvirial \citep{Parker16b,Parker18b}. These compact, highly substructured initial conditions are also the optimal initial conditions identified from the RW/WW analysis in \citet{Schoettler21b}.

\begin{table*}
  \caption[bf]{Summary of the results from the $N$-body simulations. The columns show the simulation ID, the subcluster, and then whether each quantity from the simulations is consistent with the observed values. The first of these columns is whether the numbers of runaway/walkaway stars (RW/WW) in \citet{Schoettler21b} match the observed values, then the subsequent columns are the various structural parameters we calculate in this paper. Good fits to the observations are indicated by a `Y', whereas marginal fits are indicated by an `?' and bad fits are indicated by an `N'. The final columns are the total number of good fits, (a) just considering structure $n_{\rm fit}$ (structure only), and (b) with the runaway/walkaway constraints from \citet[][$n_{\rm fit}$ with RW/WW]{Schoettler21b}.}
  \begin{center}
    \begin{tabular}{|c|c|c|c|c|c|c|c|c|}
      \hline
Sim. ID & Subcluster & RW/WW & $\tilde{\Sigma}$ & $\mathcal{Q}$ & $\mathcal{Q} - \Lambda_{\rm MSR}$ &   $\mathcal{Q} - \Sigma_{\rm LDR}$ & $n_{\rm fit}$ (structure only) & $n_{\rm fit}$ with RW/WW \\
             \hline
16-03-1 & S~Mon & {\bf Y} & Y & Y & Y & ? & 3+? & 4+? \\
16-03-1 & IRS\,1/2 & {\bf Y}  & Y & Y & Y & ? & 3+? & 4+?\\
\hline
16-05-1 & S~Mon & {\bf Y}  & Y & Y & Y & ? & 3+? & 4+? \\
16-05-1 & IRS\,1/2 & {\bf Y}  & Y & Y & Y & N & 3 & 4 \\
\hline
30-03-1 & S~Mon & {\bf N} & N & Y & N & Y & 2 & 2 \\
30-03-1 & IRS\,1/2 &  {\bf N}  & N & Y & Y & Y & 3 & 3 \\
\hline
30-05-1 & S~Mon &  {\bf N}  & ? & Y & N & Y & 2+? & 2+?\\
30-05-1 & IRS\,1/2 & {\bf ?} & N & Y & Y & Y & 3 & 3+? \\
\hline
16-03-5 & S~Mon &  {\bf N}  & Y & Y & Y & N & 3 & 3 \\
16-03-5 & IRS\,1/2 &  {\bf ?} & N & N & N & N & 0 & 0+?\\
\hline
16-05-5 & S~Mon &  {\bf N}  & Y & Y & N & N & 2 & 2 \\
16-05-5 & IRS\,1/2 &  {\bf N} & N & N & N & N & 0 & 0 \\
\hline
20-03-5 & S~Mon &  {\bf N}  & Y & N & N & N & 1 & 1\\
20-03-5 & IRS\,1/2 &  {\bf N} & N & N & N & N & 0 & 0\\
\hline
20-05-5 & S~Mon &  {\bf N} & Y & N & N & N & 1 & 1\\
20-05-5 & IRS\,1/2 &  {\bf N}  & N & N & N & N & 0 & 0\\
      \hline
    \end{tabular}
  \end{center}
  \label{results-table}
\end{table*}

\section{Discussion}
\label{discuss}

\subsection{Caveats and assumptions}

The main assumption behind our simulations is the idea that star-forming regions always form with substructure. Dynamical interactions almost always erase substructure, and so the amount of substructure in a star-forming region can be used as a dynamical clock \citep[e.g.][]{Scally02,Goodwin04a,Parker12d,Parker14b,DaffernPowell20}, which is the underlying principle in the structural analysis presented here. 

Both observations \citep{Andre10} and simulations \citep{Bate09} indicate that star-forming regions are clumpy and filamentary in their early stages, and this appears to translate into a substructured distribution for the pre-main sequence stars \citep{Gomez93,Larson95,Cartwright04,Cartwright09b}. However, it is not clear how appropriate the box-fractal method is for mimicking this substructure \citep[e.g.][]{Lomax18}. 

Our simulations are also devoid of any remaining gas from the star formation process, which could change the dynamical evolution of the subclusters if rapidly expelled by feedback from the massive stars \citep[e.g.][]{Tutukov78,Goodwin97,Baumgardt07,Shukirgaliyev18,Dinnbier20}.

In NGC\,2264 there appear to have been multiple episodes of star formation, with the S~Mon subcluster appearing to be older than IRS\,1/2. We have analysed the same $N$-body simulations and applied the results to both subclusters (albeit at their respective ages). A more realistic approach would be to model the entire NGC\,2264 region self-consistently, although such a simulation would need to be fine-tuned to account for the different bursts of star formation.

We emphasise again that we have elected to split NGC\,2264 into two separate subclusters, based on the different ages of the subclusters in question. \citet{Hetem19} analysed the overall structure of the entire NGC\,2264 region and determined the $\mathcal{Q}$-pararameter, $\Lambda_{\rm MSR}$ and $\Sigma_{\rm LDR}$ ratios for all of the stars. This means that we cannot directly compare our results to theirs. However, our values are broadly consistent with those derived by \citet{Hetem19} for the entire NGC\,2264 region, with minimal evidence for a different spatial distribution for the most massive stars. 

\subsection{The initial conditions of star formation in NGC\,2264}

Based on the numbers of runaways and walkways stars in \emph{Gaia} DR2 and \emph{Gaia} EDR3, \citet{Schoettler21b} postulated that the initial density of both the S~Mon and IRS~1/2 subclusters in NGC\,2264 was likely to be of order $10^4$\,M$_\odot$\,pc$^{-3}$, and furthermore that this star-forming region was consistent with substructured and subvirial initial conditions.

Our structural analyses is also consistent with these initial conditions, with the possible exception of the $\mathcal{Q} - \Sigma_{\rm LDR}$ plots, which could indicate a marginally less dense set of initial conditions ($\geq 1000$\,M$_\odot$\,pc$^{-3}$). This is because in our simulations, stars that have spatially correlated velocities tend to group around the most massive stars, leading to elevated relative local surface densities for the most massive stars. However, if a dataset is incomplete, or affected by extinction, the $\Sigma_{\rm LDR}$ ratio could be reduced. 

In terms of other nearby star-forming regions on which similar analyses have been performed, NGC\,2264 appears to have had much more dense initial conditions than IC\,348 \citep{Parker17a}, $\rho$~Oph \citep{Parker12c,Parker14b} and Cyg~OB2 \citep{Wright14}, but comparable initial conditions to the ONC \citep{Parker14b,Schoettler20} and NGC\,1333 \citep{Parker17a}. 

Although just a handful of star-forming regions, these results indicate that nearby star-forming regions do not have a common initial stellar density, but rather a range of initial densities. At the high end, regions with densities $\sim 10^4$\,M$_\odot$\,pc$^{-3}$ could truncate protoplanetary discs via direct interactions \citep{Scally02,Vincke15,Vincke16,Zwart16,Winter18a}, although significant destruction/alteration of the disc population via truncation only occurs if these densities are maintained for several Myr \citep{Vincke18,Winter18b}. 

In our simulations that reproduce the observed numbers of runaway and walkaway stars from NGC\,2264, and match the observed surface density, structure and levels of mass segregation, the initial average volume densities are high ($\tilde{\rho} = 10^4$\,M$_\odot$\,pc$^{-3}$), but rapidly decrease to $\tilde{\rho} = 100$\,M$_\odot$\,pc$^{-3}$. In this scenario, some initial disc truncation could occur, but this is likely to be short-lived.

However, if planets have already been able to form at these young ages \citep[which recent observations appear to confirm, e.g.][]{Alves20,SeguraCox20}, they are likely to have their orbits affected at densities in the range $100 - 1000$\,M$_\odot$\,pc$^{-3}$ \citep{Adams06,Parker12a}. At the lower end of the scale, regions with densities $<100$\,M$_\odot$\,pc$^{-3}$ could still affect planet formation if massive stars are present due to Far Ultra Violet (FUV) radiation from stars more massive than 10\,M$_\odot$, which can lead to photoevaporation of protoplanetary discs \citep{Scally02,Adams04,ConchaRamirez19,Nicholson19a,Parker21a}. Given that S~Mon and IRS~1 are both more massive than 10\,M$_\odot$ \citep[e.g.][]{Peretto06,Maiz19}, photoevaporation has almost certainly altered protoplanetary discs in NGC\,2264.

\section{Conclusions}
\label{conclude}

We have quantified the spatial structure of two subclusters within the NGC\,2264 star-forming region, S~Mon and IRS\,1/2, as well as calculating the amount of mass segregation according to $\Lambda_{\rm MSR}$ and the relative local surface densities of the massive stars, $\Sigma_{\rm LDR}$. We then compared these quantities to the output of $N$-body simulations to constrain the initial conditions for star formation in NGC\,2264. Our conclusions are the following:

(i) Both S~Mon and IRS\,1/2 have moderate $\mathcal{Q}$-parameters ($\mathcal{Q} \sim 0.8$), which indicates neither a substructured nor a very centrally concentrated distribution. Neither subcluster exhibits mass segregation according to $\Lambda_{\rm MSR}$, but the younger IRS\,1/2 subcluster does exhibit a significantly high local surface density ratio for the most massive stars ($\Sigma_{\rm LDR} > 1$).

(ii) When compared to $N$-body simulations, we find that both regions are consistent with high initial densities ($\tilde{\rho} \sim 10^4$\,M$_\odot$\,pc$^{-3}$), a high degree of initial substructure and (sub)virial initial velocities (under the assumption that star formation must always produce a filamentary/substructured distribution).

(iii) These initial conditions are also those that best match the numbers of runaway ($>30$\,km\,s$^{-1}$) and walkaway ($>5$\,km\,s$^{-1}$) stars observed with \emph{Gaia} \citep{Schoettler21b}, suggesting that analysis of the spatial structure is consistent with analysis of the proper motion velocities of stars. 

(iv) Our constraints suggest high initial densities for star formation in NGC\,2264, commensurate with similar estimates for the ONC. At these densities, protoplanetary discs and fledgling planetary systems could be greatly affected by dynamical encounters and photoevaporation from the FUV radiation fields from massive stars. 

In similar previous  analyses we have found that such high initial densities are not observed/postulated for all star-forming regions, however, providing further evidence that ongoing star formation in the Milky Way results in a range of different densities (and by extension, different initial conditions for fledgling planetary systems).

\section*{Acknowledgements}

We thank the anonymous referee for their comments and suggestions. RJP acknowledges support from the Royal Society in the form of a Dorothy Hodgkin Fellowship. CS acknowledges PhD funding from the 4IR Science and Technology Facilities Council (STFC) Centre for Doctoral Training in Data Intensive Science.

This work has made use of data from the European Space Agency (ESA) mission {\it Gaia} (\url{https://www.cosmos.esa.int/gaia}), processed by the {\it Gaia} Data Processing and Analysis Consortium (DPAC, \url{https://www.cosmos.esa.int/web/gaia/dpac/consortium}). Funding for the DPAC has been provided by national institutions, in particular the institutions participating in the {\it Gaia} Multilateral Agreement. 
This research has made use of the SIMBAD database, operated at CDS and the VizieR catalogue access tool, CDS, Strasbourg, France.

\section*{Data Availability Statement}
The data underlying this article were accessed from the \textit{Gaia} archive, \url{https://gea.esac.esa.int/archive/}. The derived data generated in this research will be shared on reasonable request to the corresponding author.

\bibliographystyle{mnras}  
\bibliography{general_ref}

\appendix

\section{Binary systems}
\label{appendix:binaries}

The structural diagnostics we apply to the observed and simulated data ($\mathcal{Q}$, $\Lambda_{\rm MSR}$ and $\Sigma_{\rm LDR}$) could in principle be biased by the binary star population. In the observational data, we have little or no information on whether each individual star is a binary, yet in order to make a meaningful assessment of the initial conditions, we require our simulations to include a population of primordial binary stars. 

\begin{figure*}
  \begin{center}
\setlength{\subfigcapskip}{10pt}
\hspace*{-0.3cm}\subfigure[10\,au cut, $\mathcal{Q}$-parameter]{\label{16-03-1_bins-a}\rotatebox{270}{\includegraphics[scale=0.27]{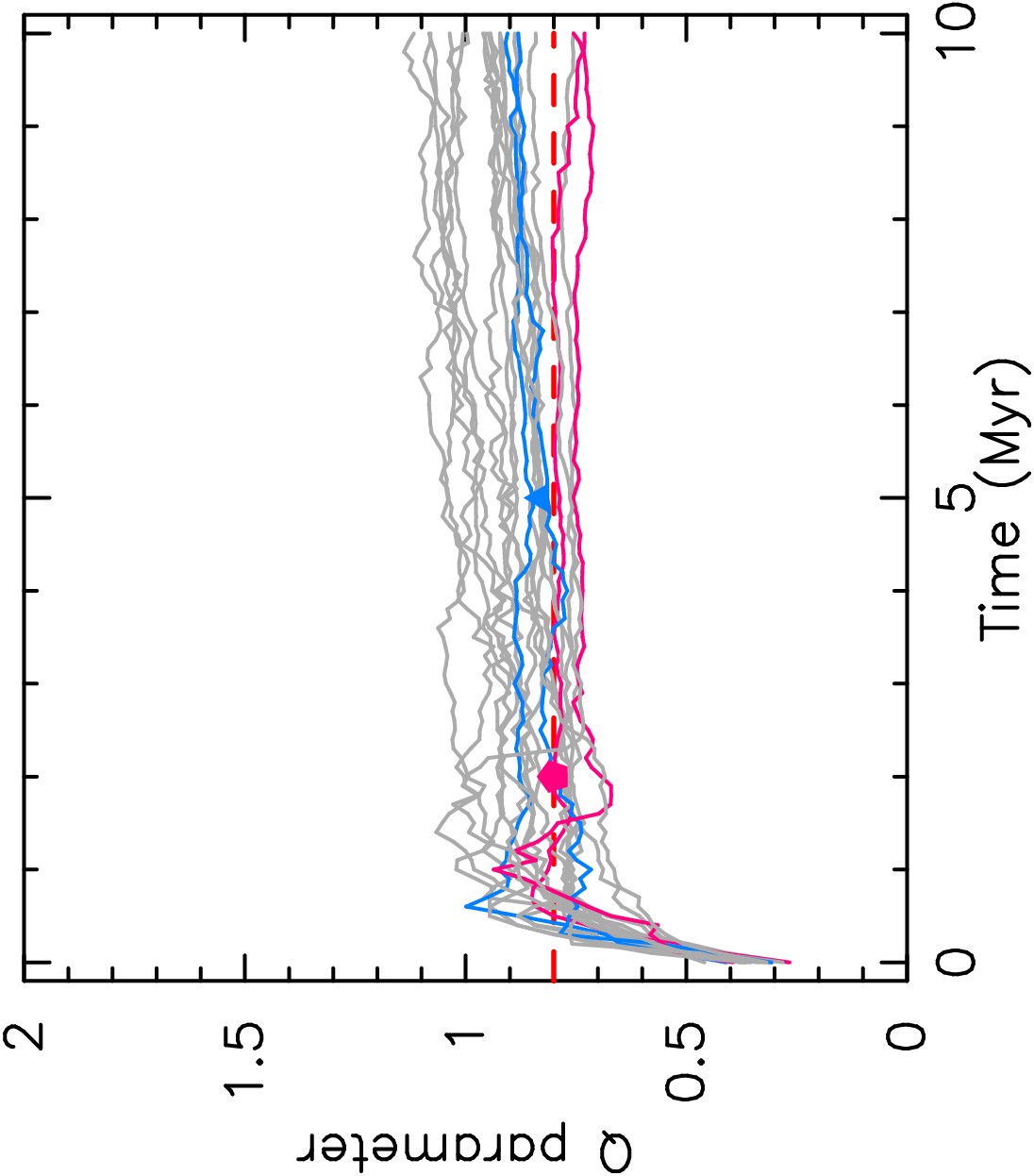}}}
\hspace*{0.1cm}\subfigure[10\,au cut, $\mathcal{Q} - \Lambda_{\rm MSR}$]{\label{16-03-1_bins-b}\rotatebox{270}{\includegraphics[scale=0.27]{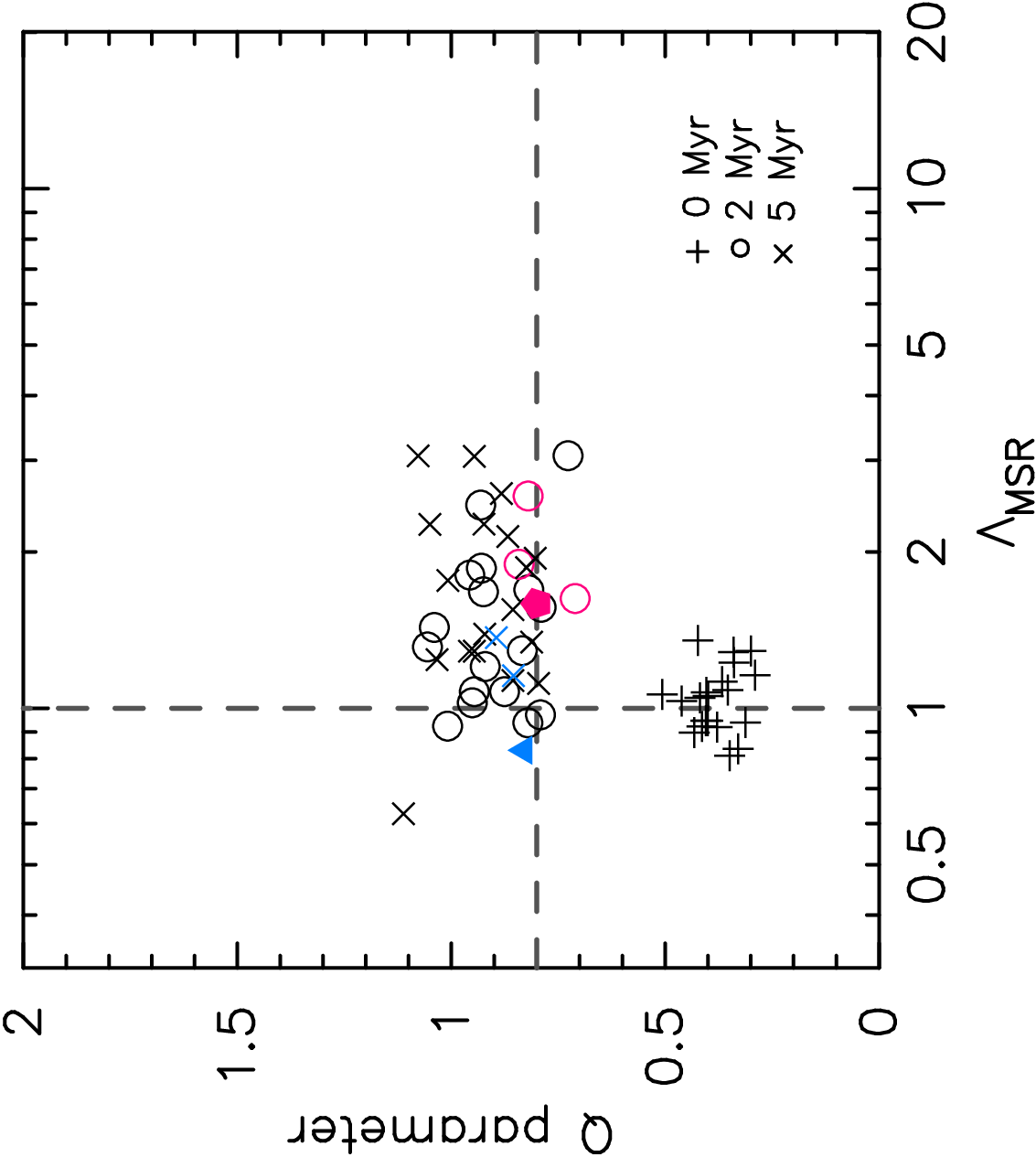}}}
\hspace*{0.1cm}\subfigure[10\,au cut, $\mathcal{Q} - \Sigma_{\rm LDR}$]{\label{16-03-1_bins-c}\rotatebox{270}{\includegraphics[scale=0.27]{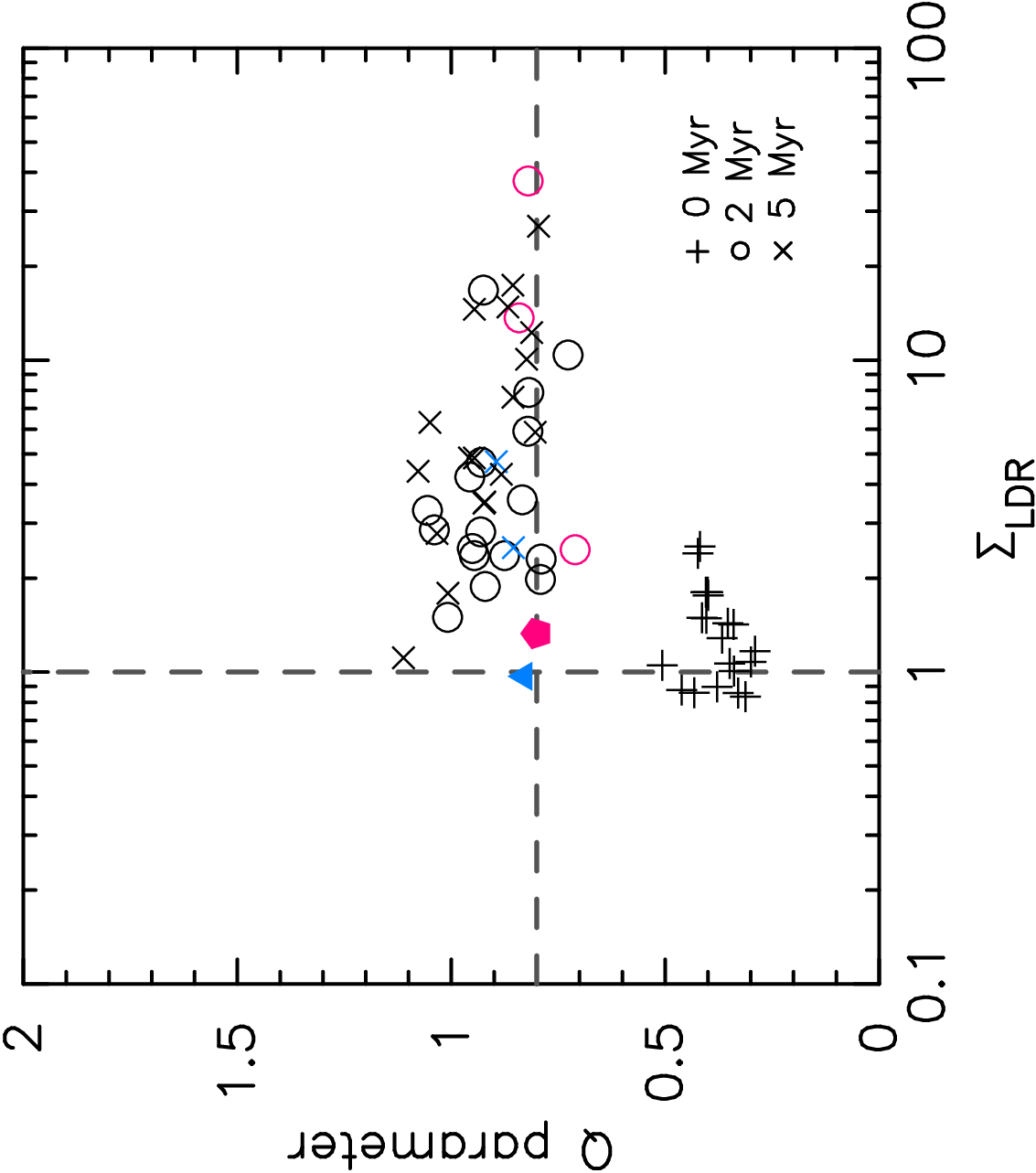}}}
\hspace*{-0.3cm}\subfigure[1000\,au cut, $\mathcal{Q}$-parameter]{\label{16-03-1_bins-d}\rotatebox{270}{\includegraphics[scale=0.27]{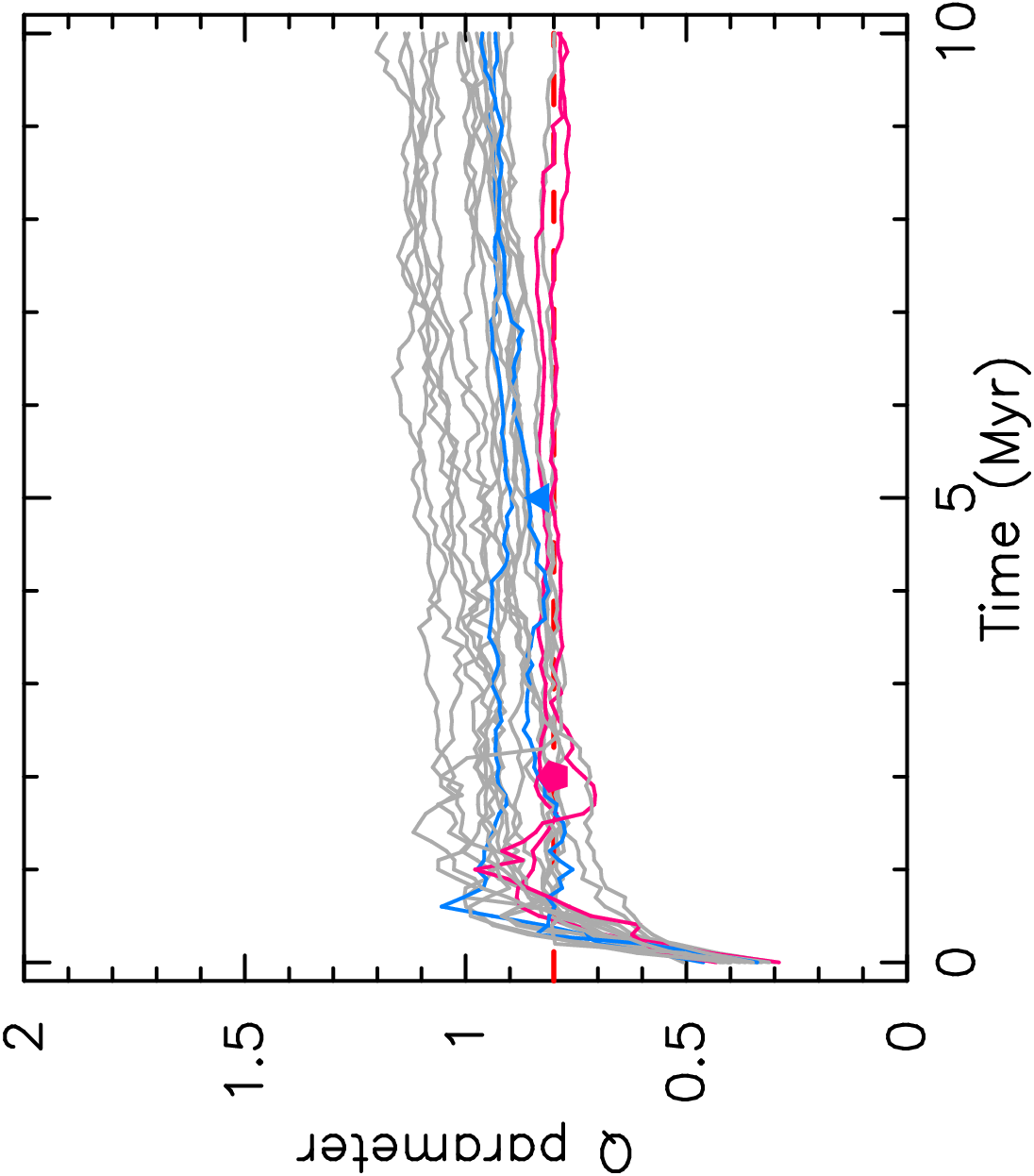}}}
\hspace*{0.1cm}\subfigure[1000\,au cut, $\mathcal{Q} - \Lambda_{\rm MSR}$]{\label{16-03-1_bins-e}\rotatebox{270}{\includegraphics[scale=0.27]{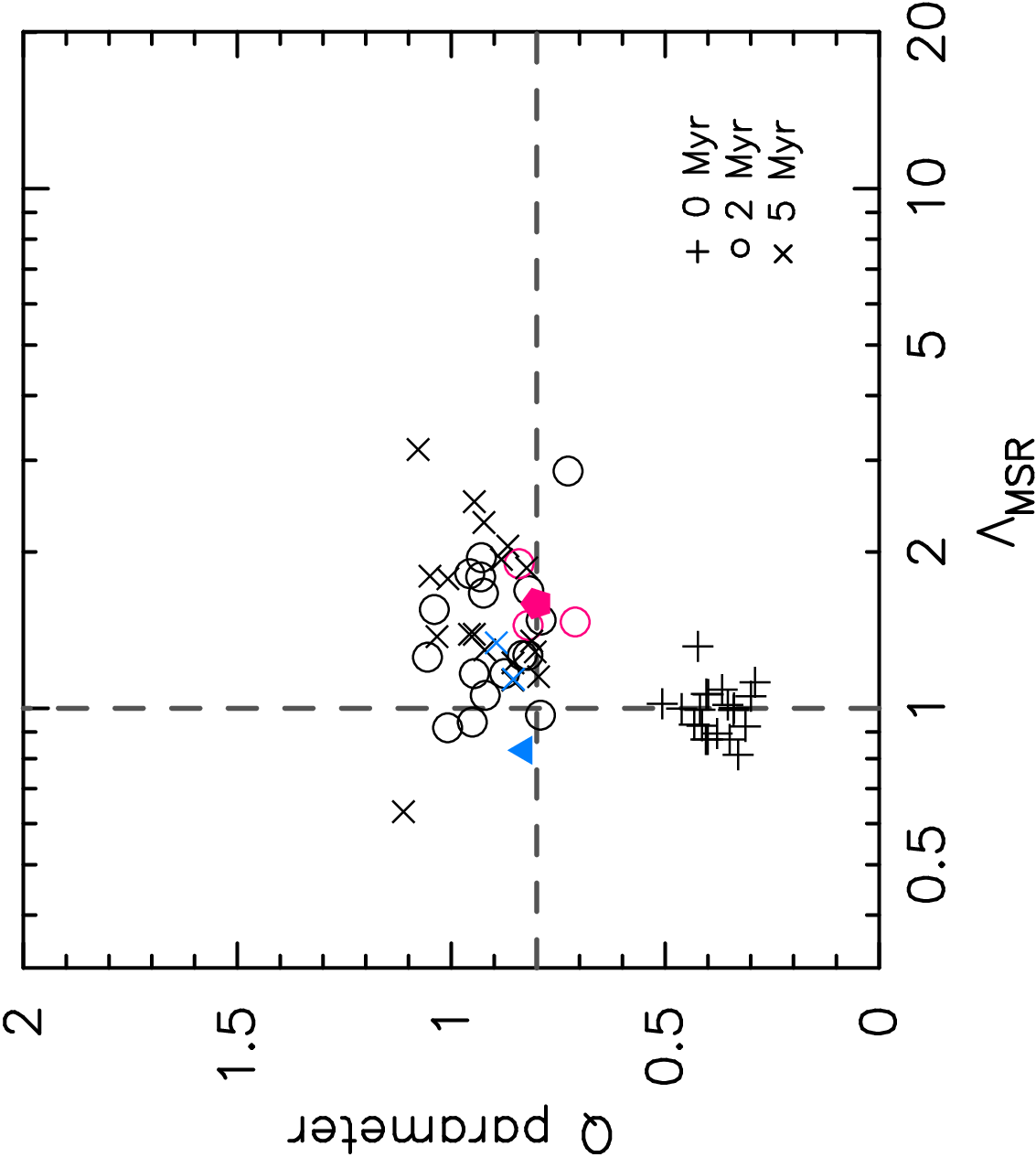}}}
\hspace*{0.1cm}\subfigure[1000\,au cut, $\mathcal{Q} - \Sigma_{\rm LDR}$]{\label{16-03-1_bins-f}\rotatebox{270}{\includegraphics[scale=0.27]{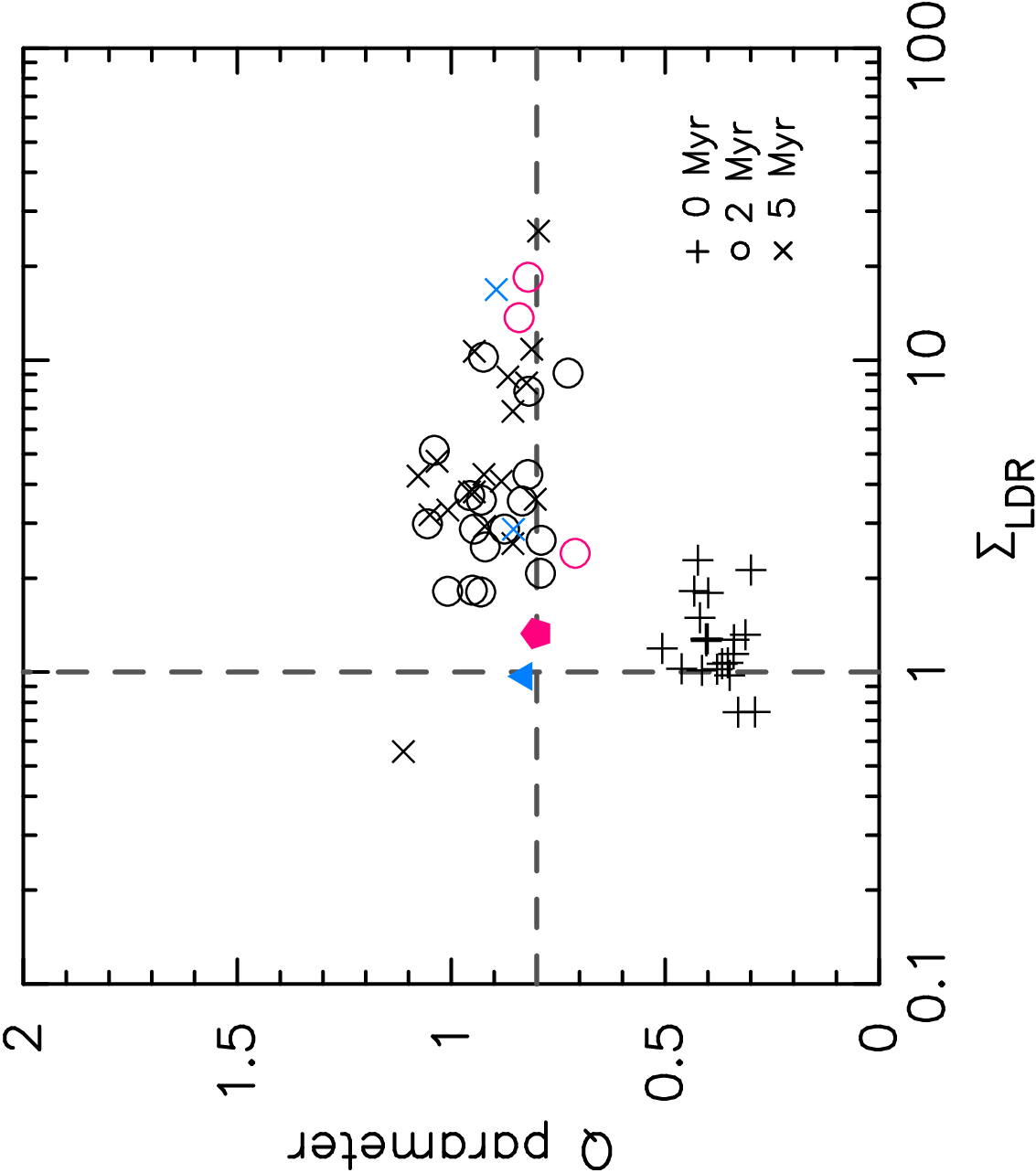}}}
\caption[bf]{More results from $N$-body simulations with highly substructured ($D = 1.6$), subvirial ($\alpha_{\rm vir} = 0.3$) initial conditions where the star-forming regions have initial radii of 1\,pc (labelled 16-03-1 in Table~\ref{simulations}, and shown in Fig.~\ref{16-03-1}). We have reanalysed these simulations by assuming binary systems with semimajor axes $a < 10$\,au would be observed as single stars (top row, panels (a)--(c)) and then again assuming binary systems with semimajor axes $a < 1000$\,au would be observed as single stars (bottom row, panels (d)--(f)).  In all panels the pink pentagon corresponds to the observed values for the IRS\,1/2 subcluster and the blue triangle corresponds to the observed values for the S~Mon subcluster. The pink lines and symbols are simulations where the numbers of runaways and walkaways are consistent with the observed values for IRS\,1/2 at the upper age limit for this subcluster and the blue lines and symbols are simulations where the numbers of runaways and walkaways are consistent with the observed values for S~Mon at the upper age limit for this subcluster. Panels (a) and (d) show the evolution of the $\mathcal{Q}$-parameter [the horizontal red dashed line indicates the boundary between substructured ($\mathcal{Q}<0.8$) and smooth ($\mathcal{Q}>0.8$) morphologies].  Panels (b) and (e) show the $\mathcal{Q}$-parameter and $\Lambda_{\rm MSR}$ values at 0, 2 and 5\,Myr and panels (c) and (f) show the $\mathcal{Q}$-parameter and $\Sigma_{\rm LDR}$ values at 0, 2 and 5\,Myr. The horizontal lines in panels (b), (c), (e)  and (f) show the boundary between substructured and smooth morphologies according to the $\mathcal{Q}$-parameter, and the vertical lines indicate $\Lambda_{\rm MSR} = 1$ (panels (b) and (e)) and $\Sigma_{\rm LDR} = 1$ (panels (c) and (f)).} 
\label{16-03-1_bins}
  \end{center}
\end{figure*}

\citet{Kupper11} show that the $\mathcal{Q}$-parameter is slightly sensitive the the adopted binary population, but both $\Lambda_{\rm MSR}$ and $\Sigma_{\rm LDR}$ have not been tested in distributions that contain binaries. Moreover, because the binary population in NGC\,2264 is largely unknown, we have reproduced the analysis of our best-fitting simulation ($r_F = 1$\,pc, $D = 1.6$, $\alpha_{\rm vir} = 0.3$, Fig.~\ref{16-03-1}) under the assumption that we can resolve the individual components of binaries with separations greater than 10\,au (panels (a)--(c) in Figs.~\ref{16-03-1_bins}) or 1000\,au (panels (d)--(f) in Fig.~\ref{16-03-1_bins}). If a binary has a separation less than one of these thresholds, we count the system as a single star (in terms of its mass, and position), whereas if the binary separation exceeds these thresholds, we treat the components as two separate stars.

Fig.~\ref{16-03-1_bins} clearly shows that selecting a separation threshold for binaries makes minimal differences to the outcome of the analysis and so we simply run our analysis on the population of stars within twice the half-mass radius from the subcluster centres, as described in Section~\ref{nbody_method}.

\label{lastpage}

\end{document}